\documentclass{jfm}

\renewcommand{\thesection}{\arabic{section}}
\renewcommand{\thesubsection}{\thesection.\arabic{subsection}}
\renewcommand{\thesubsubsection}{\thesubsection.\arabic{subsubsection}}

\usepackage[a4paper, total={6.5in, 9.5in}]{geometry}
\usepackage{graphicx}
\usepackage{dcolumn}
\usepackage{bm}
\usepackage{epstopdf, epsfig}

\usepackage{mathrsfs}
\usepackage[round,sort]{natbib}
\usepackage{amsmath}
\usepackage{amssymb}
\usepackage{verbatim}
\usepackage{ctable}
\usepackage{color}
\usepackage{soul}
\usepackage[normalem]{ulem}
\newcommand{\stkout}[1]{\ifmmode\text{\sout{\ensuremath{#1}}}\else\sout{#1}\fi}

\DeclareFontFamily{OT1}{pzc}{}
\DeclareFontShape{OT1}{pzc}{m}{it}{<-> s * [1.10] pzcmi7t}{}
\DeclareMathAlphabet{\mathpzc}{OT1}{pzc}{m}{it}

\newcommand{\Webe}{\operatorname{\mathit{W\kern-.20em e}}}
\newcommand{\Capp}{\operatorname{\mathit{C\kern-.11em a}}}

\usepackage{hyperref}

\newcommand{\q}[2]
{
\frac{\partial #1}{\partial #2}
}
\newcommand{\dd}[2]
{
\frac{\mathrm{d} #1}{\mathrm{d} #2}
}

\newcommand{\od}[1]
{
	O\left(\epsilon^{#1}\right)
}
\newcommand{\ud}[0]
{
	\,\mathrm{d}
}

\newcommand{\ar}[0]
{
\mathscr{A}
}

\newcommand{\csat}[0]
{
{c_\mathrm{sat}}
}

\newcommand{\tit}{\textit}
\newcommand{\mbf}{\mathbf}

\newcommand{\br}{\bar{r}}

\newcommand{\Ca}{\mbox{Ca}}
\newcommand{\Bo}{\mbox{Bo}}

\definecolor{amber}{rgb}{1.0, 0.75, 0.0}

\begin{document}

\shorttitle{
Evaporation of non-circular droplets
}
\shortauthor{A. W. Wray and M. R. Moore}
\title{Evaporation of non-circular droplets}

\author{Alexander W. Wray\aff{1}
  \corresp{\email{alexander.wray@strath.ac.uk}},
 \and Matthew R. Moore\aff{2}}

\affiliation{\aff{1}Department of Mathematics and Statistics, University of Strathclyde, Livingstone Tower,
26 Richmond Street, Glasgow G1 1XH, UK
\aff{2}Department of Mathematics, School of Natural Sciences, University of Hull, Cottingham Road, Hull, HU6 7RX, UK}

\maketitle

\begin{abstract}
The dynamics of thin, non-circular droplets evaporating in the diffusion-limited regime are examined. The challenging non-rectilinear mixed-boundary problem this poses is solved using a novel asymptotic approach and an asymptotic expansion for the evaporative flux from the free surface of the droplet is found. While theoretically valid only for droplets that are close to circular, it is demonstrated that the methodology can successfully be applied to droplets with a wide variety of footprint shapes, including polygons and highly non-convex domains. While the applications of this are numerous, the analytically-tractable case of deposition of solute from large droplets is examined in detail, including a matched asymptotic analysis to resolve the pressure, streamlines and deposition up to second order.
\end{abstract}


\section{Introduction}
\label{sec:Introduction}

Evaporation of droplets is ubiquitous in the world around us. However, despite the apparent simplicity of the geometry, the dynamics involved are typically very complex. Gaining a theoretical understanding of the process is thus of particular importance due to the key role droplet evaporation plays in everything from inkjet printing, to the spreading of pesticides on leaves, to diagnostic applications of blood drying \cite[][]{brutin2018recent, mampallil2018review}. As a result, determining the evaporative flux of liquid from the free surface into the surrounding gas has become a key goal of modellers, as this will drive the internal fluid motion, and consequential dynamics of the droplet.

However, even within well-established regimes such as \tit{diffusion-limited} evaporation, finding analytical expressions for the evaporative flux is difficult, while numerical simulations can be expensive and challenging. In diffusion-limited evaporation, vapour diffuses away from the free surface of the droplet sufficiently quickly that the process may be taken to a reasonable approximation to be steady, so that the concentration of vapour satisfies a mixed boundary value problem for Laplace's equation. Mathematically, this is a notoriously challenging problem, with singularities induced at the contact line of the droplet. Perhaps unsurprisingly, the solution --- and hence, the flux --- depends strongly on the geometry of the droplet. Exact solutions are scarce: some examples of known solutions for the evaporative flux of droplets in isolation include the flat disk \citep{weber1873ueber}, flat ellipse \citep{boersma1993solution}, spherical cap \citep{popov2005evaporative} and ellipsoidal cap \citep{Kellogg1929} but few others.

For more complicated geometries where techniques such as separation of variables and transform methods fail, we can make progress when the droplet profile is such that it may reasonably be treated as \tit{thin}. For many situations, once a droplet has been deposited onto the substrate, its contact line becomes pinned on surface roughnesses so that the thin assumption is equivalent to assuming that a typical contact radius, say $R$, is much larger than the initial thickness of the droplet, say $H$, so that $H/R\ll1$. Pinning persists for the majority of the evaporation \cite[][]{Hu2002} and, indeed, may be further enhanced when solute accumulates at the edge of the droplet \cite[see, for example,][]{orejon2011stick,weon2013self}, so that the thin assumption continues to hold. 

When a droplet is thin, the Laplace problem may be linearised into a half-space problem, so that the most sensible starting point for finding the evaporative flux is to use a Green's function formulation to relate the evaporative flux on the droplet to the known vapour saturation concentration through an integral equation. Various different approaches for expanding the Green's function kernel may then be used to invert the integral. For simpler geometries, the solution may be found exactly, such as a disk \citep{copson1947problem,sneddon1966mixed} or --- for suitable saturation concentrations --- for an ellipse \citep{galin1961contact}. However, for more complicated geometries, we must again turn to numerical and approximate solutions \cite[see, for example,][]{borodachev1974contact, okon1979capacitance}.

This is particularly problematic due to the numerous uses of these evaporative models, such as in the explanation of the  famous `coffee-ring' effect \cite[][]{deegan1997capillary, deegan2000contact}: 
in many practical situations, the liquid contains a non-volatile component and the pertinent quantity of interest is the final distribution of this component once the liquid has fully evaporated. Common examples of applications include colloidal patterning and the fabrication of microscale electronics \cite[see, for example, ][]{Harris2007, Choi2010}; fabrication techniques using inkjet printing \cite[][]{Layani2009} including printing OLEDs \cite[][]{eales2015evaporation}; optical mapping of DNA \cite[][]{Jing1998}, pesticide application \cite[][]{basi2013effects} and blood analysis \cite[see][and the references therein]{smith2018wetting}. 
The final distribution is intrinsically linked to the dominant evaporative process \cite[][]{murisic2011evaporation}, which in turn depends on material and thermodynamic properties of the liquid, the surface the droplet lies on, and the surrounding atmosphere. This has resulted in a rapidly-moving field examining numerous extensions to this problem, including to residues from droplets on inclines \citep{du2015ring}, multiple droplets in proximity \citep{wray2021contact}, diffusive effects \citep{moore2021nascent, moore2022nascent}, jamming effects \citep{popov2005evaporative} and numerous attempts at control \citep{mampallil2018review}. 
However, a critical avenue is the behaviour of non-circular droplets, with the only existing studies being numerical \citep{FreedBrown2015, saenz2017dynamics} or considering the early stages of deposit formation \citep{moore2022nascent}. This is particularly surprising given the ubiquity of square/rectangular \citep{mai200753} and hexagonal \citep{huo2020real} droplets in contexts such as the printing of AMOLED screens.

Here we build upon the approach of \citet{fabrikant1986capacity}, which presents an approximate solution for an arbitrary droplet geometry. In \citet{fabrikant1986capacity}, the form of the evaporative flux is prescribed and related to an expansion of the surface concentration. However, for many problems in evaporation, we actually require the opposite --- the surface concentration is the input and we seek a pointwise representation of the evaporative flux, which can then be used in analysis of the internal flow dynamics and (when applicable) the solute transport. 

In this paper, we address this deficiency. We begin by formulating a problem for nearly-circular droplets in \textsection \ref{sec:Problem_Config}, before presenting and analysing the Green's function formulation in \textsection \ref{sec:Evap_Flux}. We find an asymptotic solution for the evaporative flux valid up to second order in terms of the perturbation parameter, which we show to be in excellent agreement to full numerical simulations of the diffusion problem. In \textsection \ref{sec:Large_drops}, we utilize our results for the specific application of determining the flow dynamics and final residue for large, nearly-circular solute-laden droplets, finding predictions of the effect of geometry on the `coffee-ring' effect that are in agreement with previous studies \cite[such as][]{FreedBrown2015,saenz2017dynamics,moore2022nascent}. Finally, given potential applications in, for example, printing OLEDs, we extend our analysis to consider regular polygonal droplets in \textsection \ref{sec:Polygon_Fluxes}, and more complex shapes in \textsection \ref{sec:otherShapes}, presenting results for the evaporative flux for general droplets, as well as the internal flow and transport dynamics for large droplets. The results are again shown to be in excellent agreement with numerical simulations.

We finish by noting that the results given herein are, to our knowledge, fundamentally new results in potential theory. As a result, while we present them in the context of evaporating droplets, we anticipate that they will be of interest to researchers in areas such as nanobubbles and nanodroplets \citep{lohse2015surface}, electrical contact resistance \citep{argatov2011electrical}, thermostatics \citep{lee1994electrostatics}, flow through a porous membrane \citep{fabrikant1985potential}, and electrodynamics \citep{jackson1999classical}, among many others.

\section{Problem formulation}
\label{sec:Problem_Config}

\begin{figure}
\begin{center}
\begin{tabular}{c}
\includegraphics[width=0.8\textwidth]{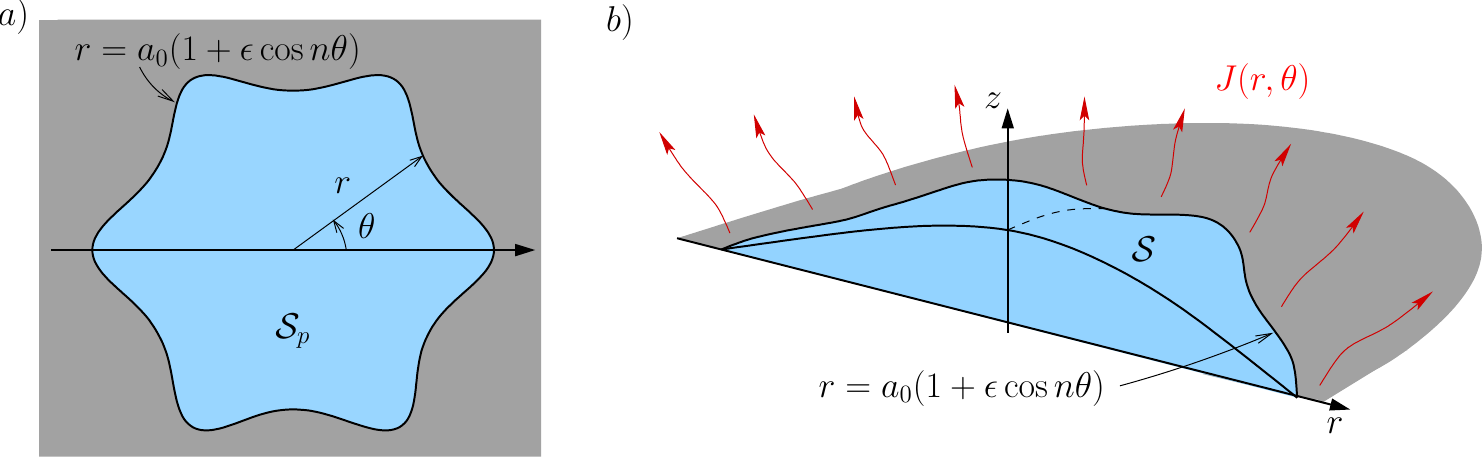}
\end{tabular}
\end{center}
\caption{A schematic showing a) a top-down and b) a side-on view of a thin, nearly-circular droplet evaporating into the surrounding atmosphere. The droplet lies on a substrate in the plane $z = 0$ and its pinned contact line is given by $r = a_{0}a(\theta) = a_{0}(1 + \epsilon\cos{n\theta})$ where $0<\epsilon\ll1$. We seek the evaporative flux of liquid vapour, $J$, from the droplet free surface, $\mathcal{S}$.}
\label{fig:Config}
\end{figure}

Consider a droplet of liquid of constant density $\rho$ and surface tension $\gamma$ lying on a flat substrate. We work in cylindrical polar coordinates $(r,\theta,z)$ with the substrate lying in the plane $z=0$. The contact line of the droplet is located at $r=a_{0}a(\theta)$, where $a_{0}$ is the mean radius. The surface of the droplet is denoted by $\mathcal{S}$. The configuration is shown schematically in Figure \ref{fig:Config}. 

We assume that the droplet is thin so that, to leading order in the droplet aspect ratio $\delta = h_{0}/a_{0}$, where $h_{0}$ is the maximum initial thickness of the droplet, it appears to be flat \citep{dunn2008mathematical}. We shall also assume that evaporation is taking place in the diffusion-limited regime \citep{sultan2005evaporation}, so that the local mass flux from the droplet $J$ may be calculated from the vapour concentration $c$ \citep{popov2005evaporative}. In the far field, $c$ approaches the constant value $c_\infty$, while on the surface of the droplet it takes the constant saturation value $\csat$. 

The system is nondimensionalised according to
\begin{equation}
    c=c_\infty+(\csat-c_\infty)\hat{c}, \quad J=\frac{D\left(\csat-c_\infty\right)\hat{J}}{a_0}, \quad (r,z)=a_0(\hat{r},\hat{z}),
\end{equation}
where $D$ denotes the diffusion coefficient of vapour in the atmosphere and carets denote dimensionless quantities. Immediately dropping the caret notation, the vapour concentration satisfies Laplace's equation
\begin{equation}
    \nabla^2c=0 \quad \mbox{in} \quad z>0,
    \label{eq:Laplace}
\end{equation}
subject to appropriate boundary conditions on $z=0$, namely
\begin{equation}
    c=1 \text{ on }\mathcal{S}_{p}\text{;}\quad \frac{\partial c}{\partial z}=0 \text{ outside }\mathcal{S}_{p},
    \label{eqn:Surface_Conds}
\end{equation}
where $\mathcal{S}_{p}$ is the projection of $\mathcal{S}$ onto the plane $z = 0$, and the far-field condition
\begin{equation}
    c\to0\quad \text{as}\quad \sqrt{r^2+z^2}\to\infty.
    \label{eq:Far_Field}
\end{equation}
Once $c$ has been determined, the flux $J(r,\theta)$ is given by 
\begin{equation}
    J =-\left.\q{c}{z}\right|_{z = 0}.
    \label{eq:Evap_Flux}
\end{equation}
for $(r,\theta)\in\mathcal{S}_{p}$.

\section{Evaporative flux}
\label{sec:Evap_Flux}
This problem may be posed in a Green's function formulation as
\begin{equation}
    c(r,\theta,z) =\frac{1}{2\pi}\iint_{\mathcal{S}_{p}} 
    G(r',\theta',0;r,\theta,z)J(r',\theta')r'\ud r'\ud \theta'.\label{eq:gfForm_full}
\end{equation}
where the standard Green's function for Laplace's equation with a Neumann condition on $z' = 0$ is
\begin{equation}
 G(r',\theta',z';r,\theta,z) = -\frac{1}{4\pi}\left(\frac{1}{|\mathbf{r}-\mathbf{r}'|} + \frac{1}{|\bar{\mathbf{r}}-\mathbf{r}'|}\right), \label{eq:gfExplicit}
\end{equation}
where $\mathbf{r} = (r\cos\theta,r\sin\theta,z)$, $\mathbf{r}' = (r'\cos\theta',r'\sin\theta',z')$ and $\bar{\mathbf{r}}$ is the image point of $\mathbf{r}$ in the plane $z' = 0$. 
Hence, by evaluating (\ref{eq:gfForm_full})--(\ref{eq:gfExplicit}) on the droplet free surface, we find from (\ref{eqn:Surface_Conds}) that
\begin{equation}
    1 =\frac{1}{2\pi}\iint_{\mathcal{S}_{p}} 
    \frac{J(r',\theta')}{\sqrt{r^2+r'^2-2rr'\cos(\theta-\theta')}}r'\ud r'\ud \theta'.\label{eq:gfForm}
\end{equation}
We examine the evaporation of a nearly-circular droplet with a monochromatic contact line of the from
\begin{equation}
a(\theta)=1+\epsilon \cos n\theta,
\end{equation}
where $n\in \mathbb{N}_{\geq2}$. Note that $n=1$ corresponds to a translation of the interface and so, even for the non-monochromatic shapes discussed in \textsection \ref{sec:Polygon_Fluxes}, this mode can be ignored without loss of generality by appropriate selection of the centre of the droplet. 
Following \citet{fabrikant1986capacity} and \citet{fabrikant1991mixed} we make use of the identity
\begin{align}
    \frac{1}{\sqrt{r^2+r'^2-2rr'\cos(\theta-\theta')}}=\frac{2}{\pi}\int_0^{\min(r,r')}& \frac{\ud x}{\sqrt{r^2-x^2}\sqrt{r'^2-x^2}}L\left(\frac{x^2}{rr'},\theta-\theta'\right),
\end{align}
where
\begin{equation}
    L(k,\psi)= \frac{1}{1-ke^{i\psi}}+\frac{1}{1-ke^{-i\psi}}-1, 
\end{equation}
to re-write \eqref{eq:gfForm} as
\begin{equation}
1=\frac{1}{\pi^2}\int_0^r \frac{\ud x}{\sqrt{r^2-x^2}}\int_0^{2\pi}\ud \theta'\int_x^{a(\theta')}\frac{r'\ud r'}{\sqrt{r'^2-x^2}}\left[ 1+2\sum_{j=1}^{\infty}\left( \frac{x^2}{r'r} \right)^j \cos j(\theta-\theta')  \right]J(r',\theta')\label{eq:fabInt}
\end{equation}
on $\mathcal{S}_{p}$. Note that this is formally only valid in a circle inscribed inside $r=a(\theta)$, but we shall see that the resulting solution does an excellent job of capturing the evaporative flux even outside of this circle (as is seen, for example, in the numerical validation in \textsection \ref{sec:modalValidation} and \textsection \ref{sec:Polygon_Fluxes}). This has solution
\begin{align}
    &J=\frac{2}{\pi}\frac{a}{\sqrt{a^2-r^2}}\left\{1+\epsilon f_1(r;n)\cos n\theta+\epsilon^2\left[f_{20}(r;n)+f_{22}(r;n)\cos 2n\theta \right]\right\},\label{eq:JSol}\\
    &f_1(r;n)=r^2\frac{1-r^{n-2}}{1-r^2}, \label{eq:f1}\\
    &f_{20}(r;n)=\frac{2r^4\left(2-r^{n-2} \right)-(n-2)(1-r^2)}{4 \left(1-r^2\right)^2}, \label{eq:f20}\\
    &f_{22}(r;n)=\frac{2r^4\left(1-r^{n-2}\right)+2(1+n)\left( 1-r^2\right)r^{2n}-3r^2\left(1-r^{2n} \right)}{4 \left(1-r^2\right)^2}, \label{eq:f22}
\end{align}
valid up to $O(\epsilon^{2})$, as may be verified via direct substitution into \eqref{eq:fabInt}. This result constitutes the main contribution of the present paper. It shall be shown that this allows for a wide range of newly (asymptotically) accessible shapes of droplets to be treated analytically.

The leading coefficient $2a/\pi\sqrt{a^2-r^2}$ in \eqref{eq:JSol} is the ansatz used by \citet{fabrikant1986capacity}. While formally only asymptotically accurate at leading order (and thus equivalent to $1/\sqrt{1-r^2}$ away from the contact line), this form exhibits the correct square root singularity exactly at the contact line \citep{jackson1999classical}. The numerator, and the exact form of the denominator, result in the evaporative flux being smooth at $r=0$ (if the prescribed contact line is smooth). Certain quantities, such as the integral flux, can be reasonably approximated by this leading-order solution \citep{fabrikant1986capacity,fabrikant1991mixed}. However, high accuracy spatial resolution of the flux requires the higher-order corrections given in (\ref{eq:JSol}). Note that, for example, the contours of constant evaporative flux given by $a/\sqrt{a^2-r^2}=c$ are $r=a\sqrt{1-c^{-2}}$, so that all evaporative flux contours are exactly scaled copies of the contact line, which is generally in stark contrast to the solutions given by direct numerical simulations, as we see in for various droplet geometries in Figures \ref{fig:validation}, \ref{fig:validationTriangleSquare} and \ref{fig:otherPolyResults}.

It is well known that the late-time dynamics of solute transport in the coffee-ring effect is governed by the evaporative flux about the stagnation point in the droplet interior (here, at its centre) \cite[][]{witten2009robust}. Therefore, for reference, we note that the evaporative flux given by (\ref{eq:JSol})  satisfies
\begin{align}
    J&=\frac{2}{\pi}\left[1-\frac{\epsilon^2}{4}\left( n-2 \right)\right]+O(r^2)\quad \text{as} \quad r\to 0.
    \label{eq:JStagPoint}
    \end{align}
Notably, the effect of small azimuthal variations in the droplet contact set are significantly weaker for $n = 2$ than for other modes. 

Moreover, a key factor in determining properties of the developing coffee ring in the diffusion-limited evaporative regime is the behaviour of the flux $J$ local to the contact line \cite{moore2021nascent, moore2022nascent, saenz2017dynamics}. For reference, this asymptotic behaviour of $J$ close to the contact line is given explicitly in \textsection \ref{sec:innerRegionFluxAndPressure}. For moderate values of $\epsilon$ and/or $n$, this expansion quantitatively demonstrates that the flux is enhanced close to regions of high curvature of the contact line and inhibited at regions of lower curve. However, we note briefly that $J$ can diverge (unphysically) towards negative infinity if $\epsilon$ and/or $n$ are chosen to be too large.

The total flux of liquid into the surrounding air, $F$, is given by
\begin{equation}
    F=\iint_{\mathcal{S}}J\,\ud S = 4+\epsilon^2n+O(\epsilon^4).
    \label{eq:F}
\end{equation}
We note that this result can be determined more rapidly via the reciprocal theorem \citep{fabrikant1987diffusion}.

\subsection{Validation}\label{sec:modalValidation}

\begin{figure}
\begin{center}
\begin{tabular}{cc}
\includegraphics[width=0.25\textwidth]{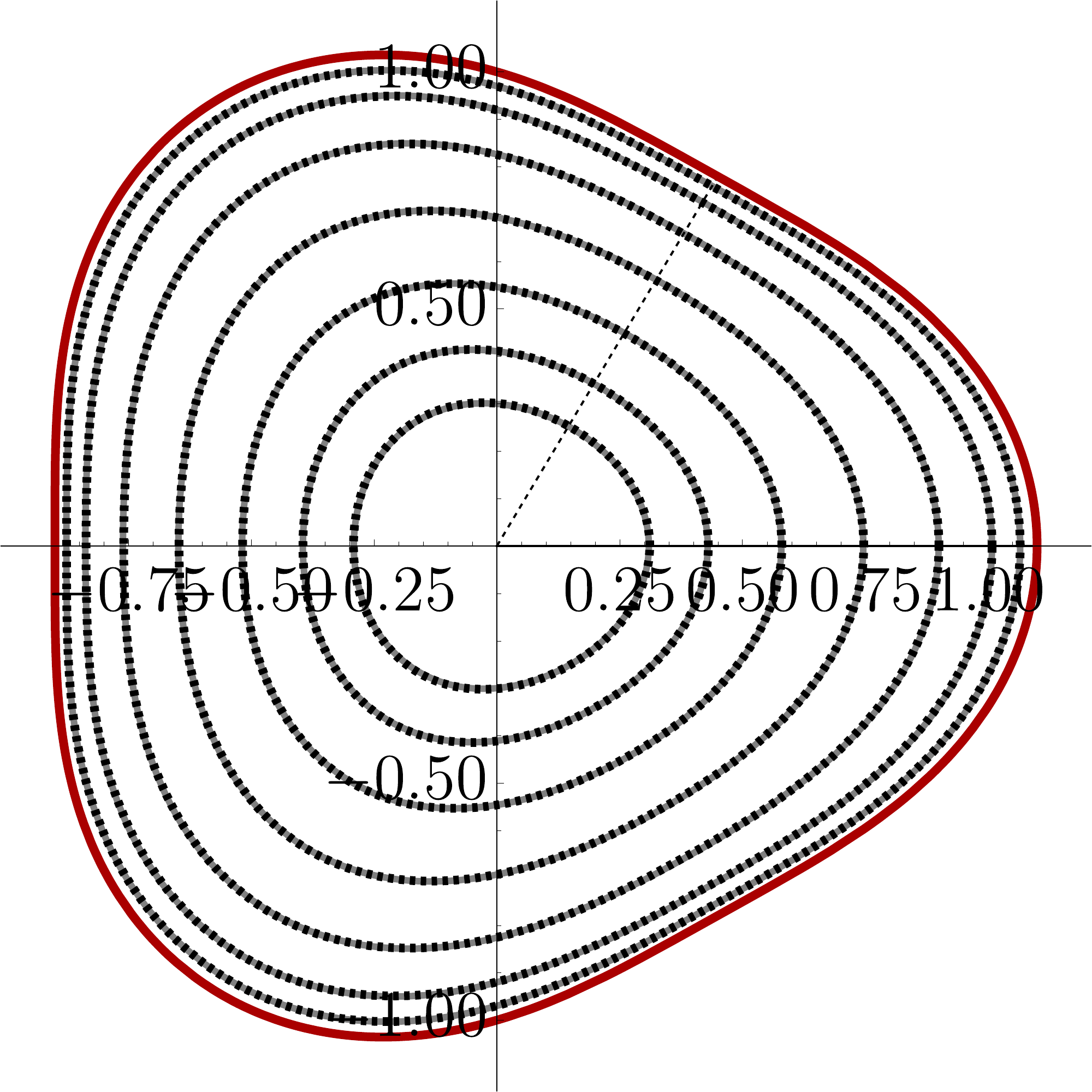}&
\includegraphics[width=0.45\textwidth]{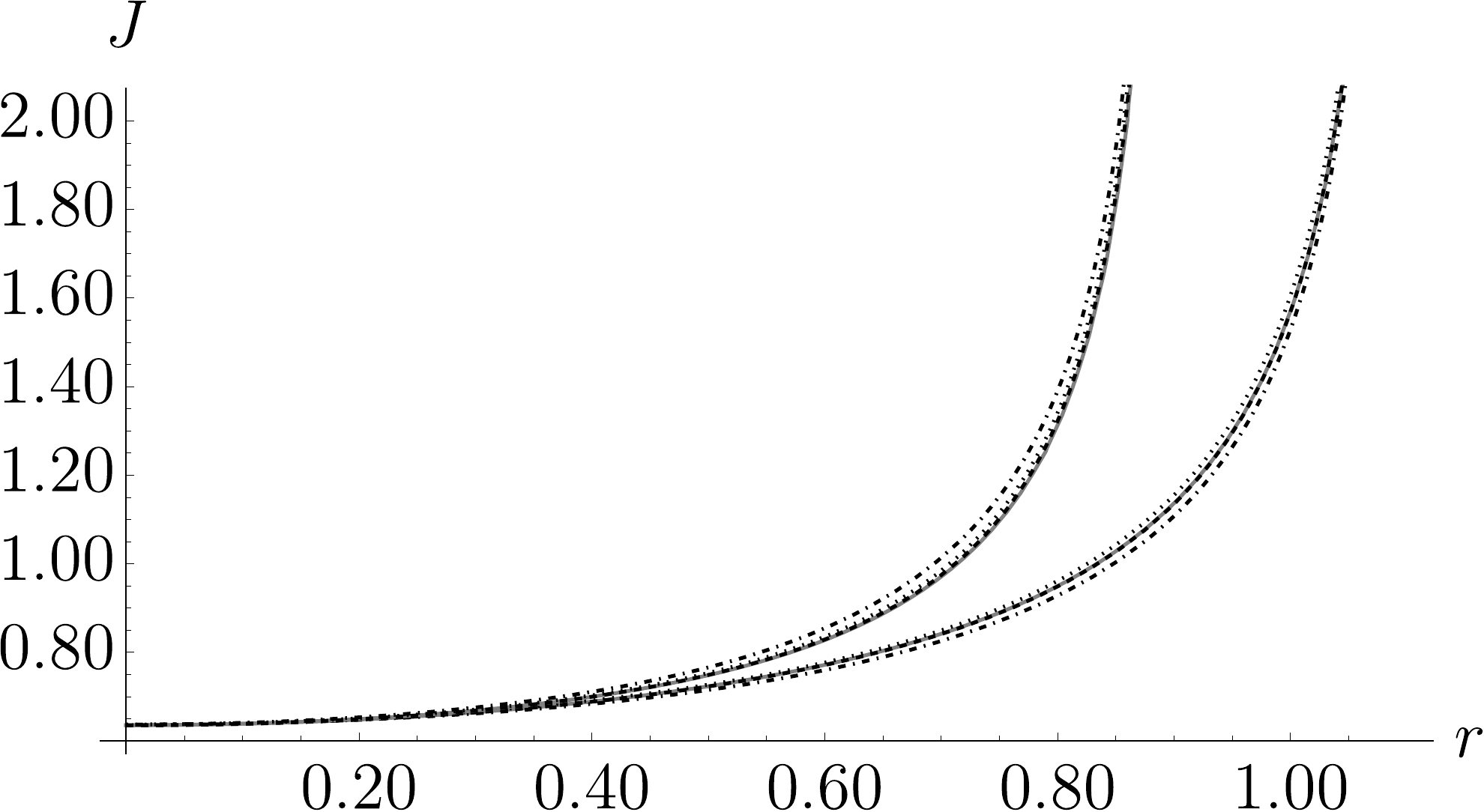}\\
(a) & (b) \\
\includegraphics[width=0.25\textwidth]{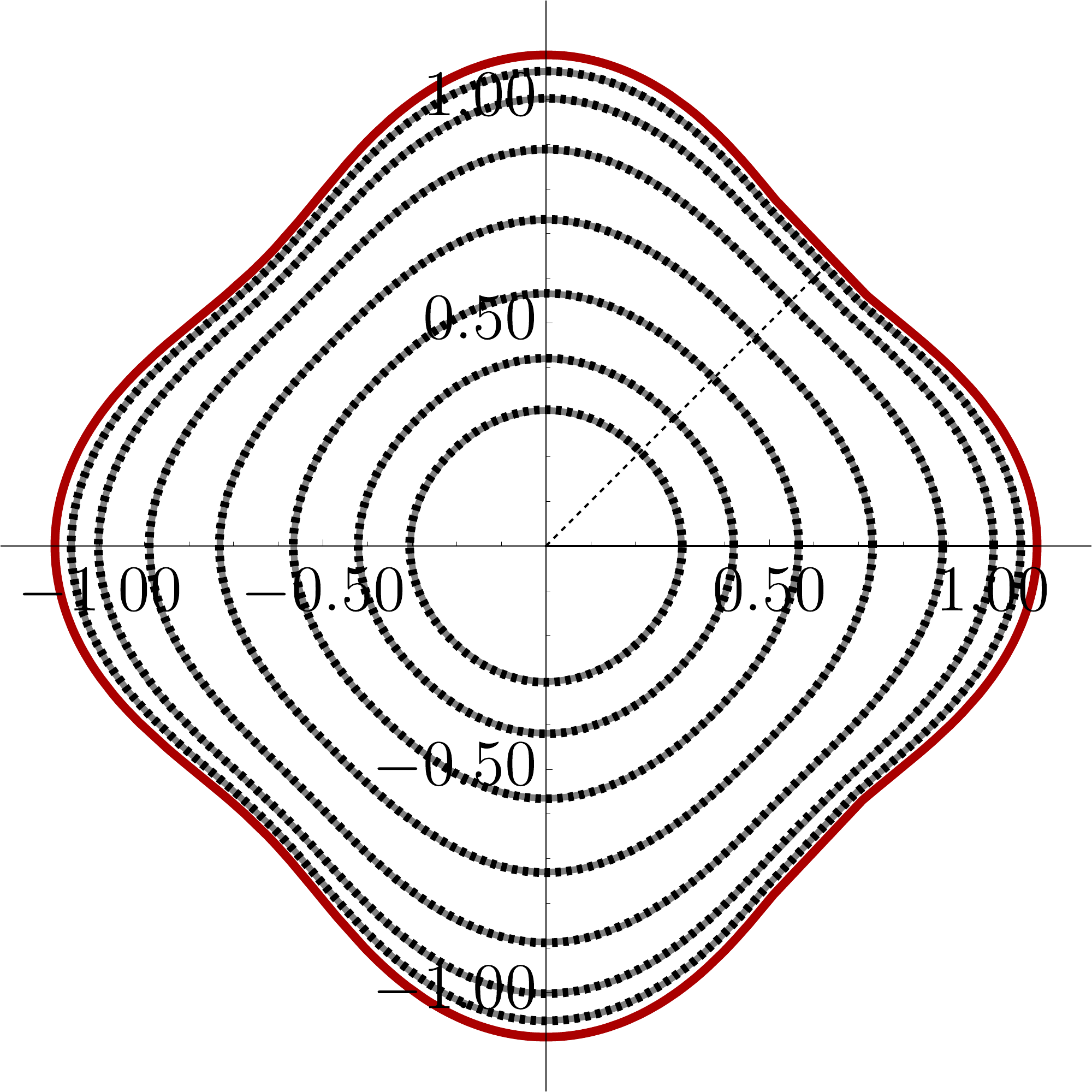}&
\includegraphics[width=0.45\textwidth]{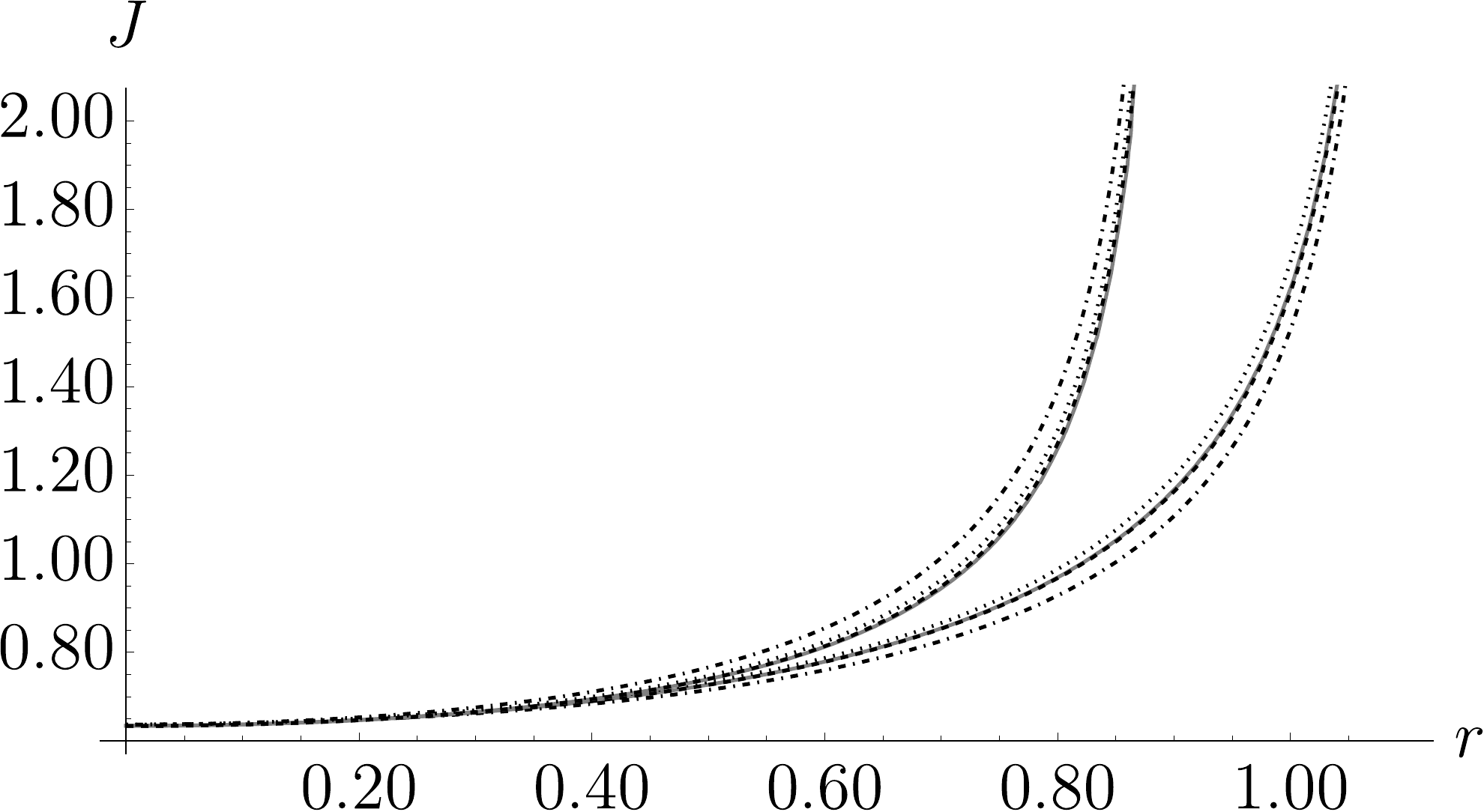}\\
(c) & (d) \\
\includegraphics[width=0.25\textwidth]{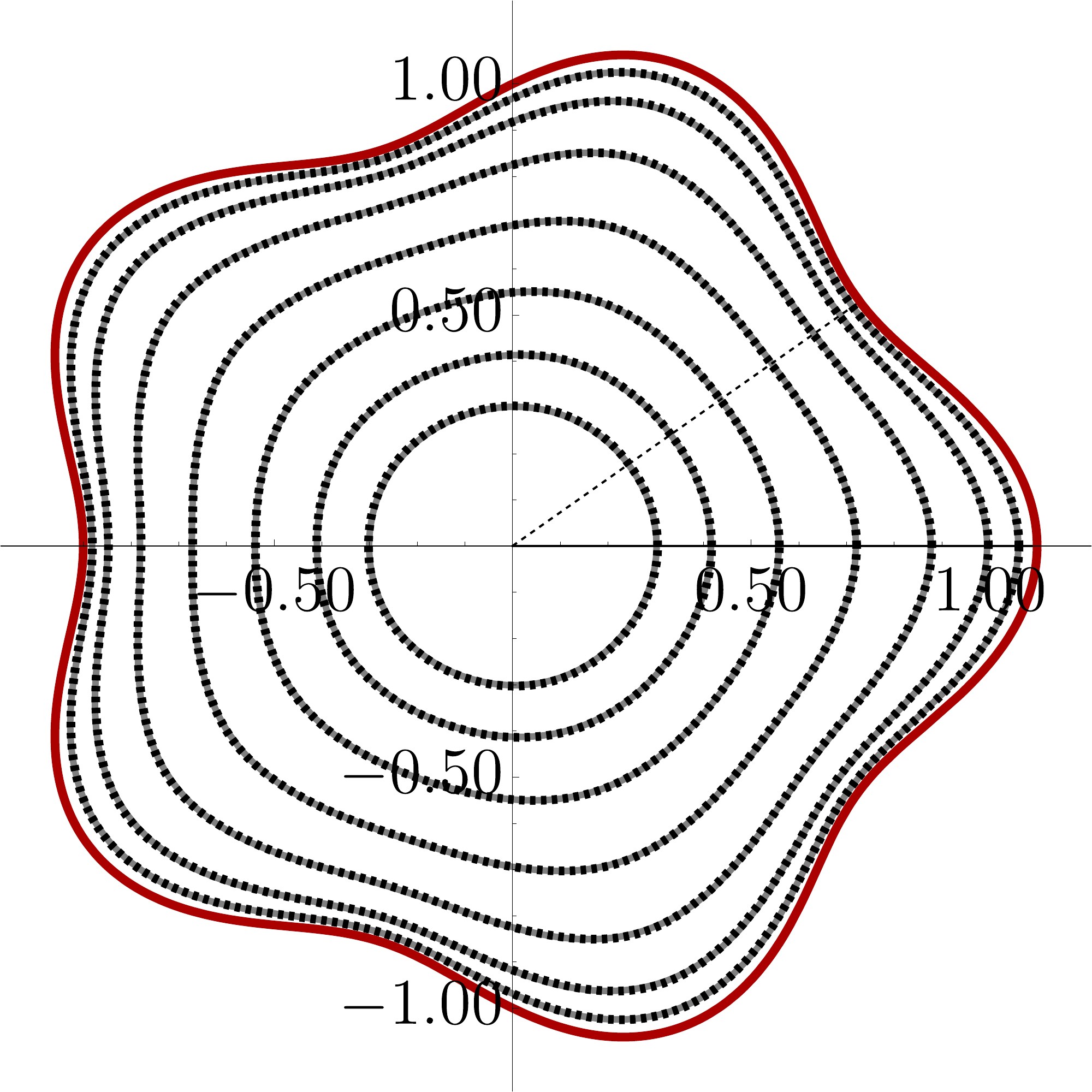}&
\includegraphics[width=0.45\textwidth]{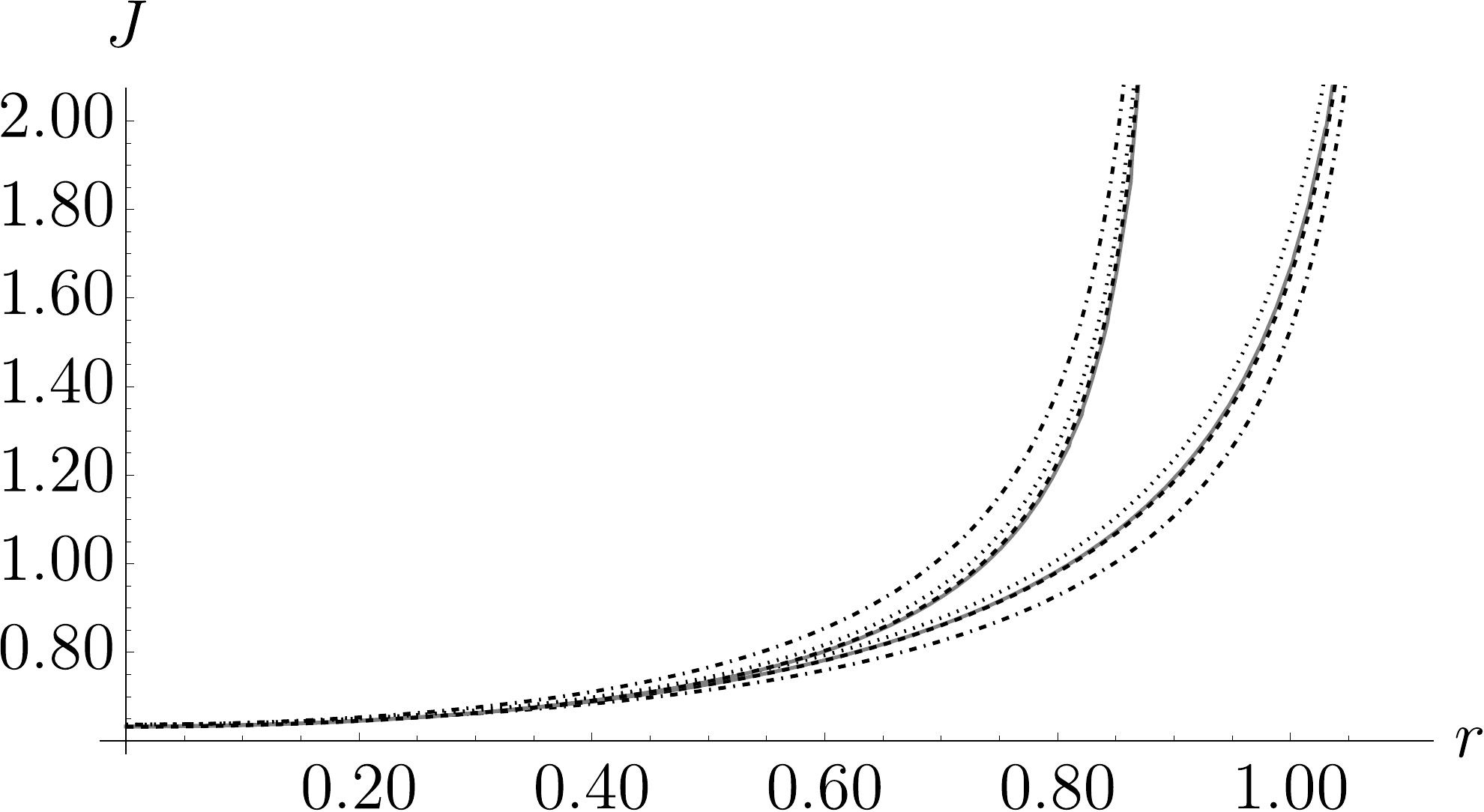}\\
(e) & (f) \\
\includegraphics[width=0.25\textwidth]{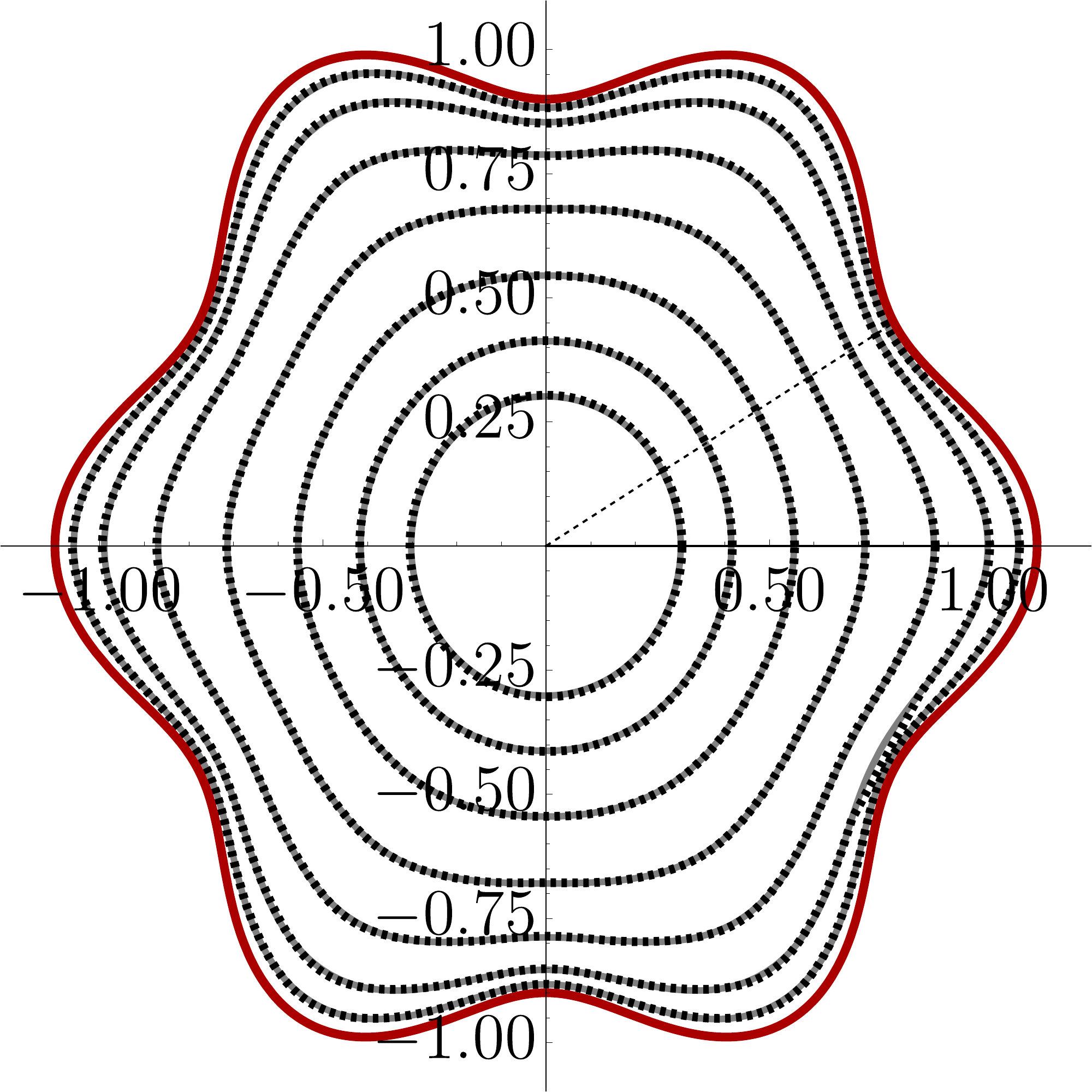}&
\includegraphics[width=0.45\textwidth]{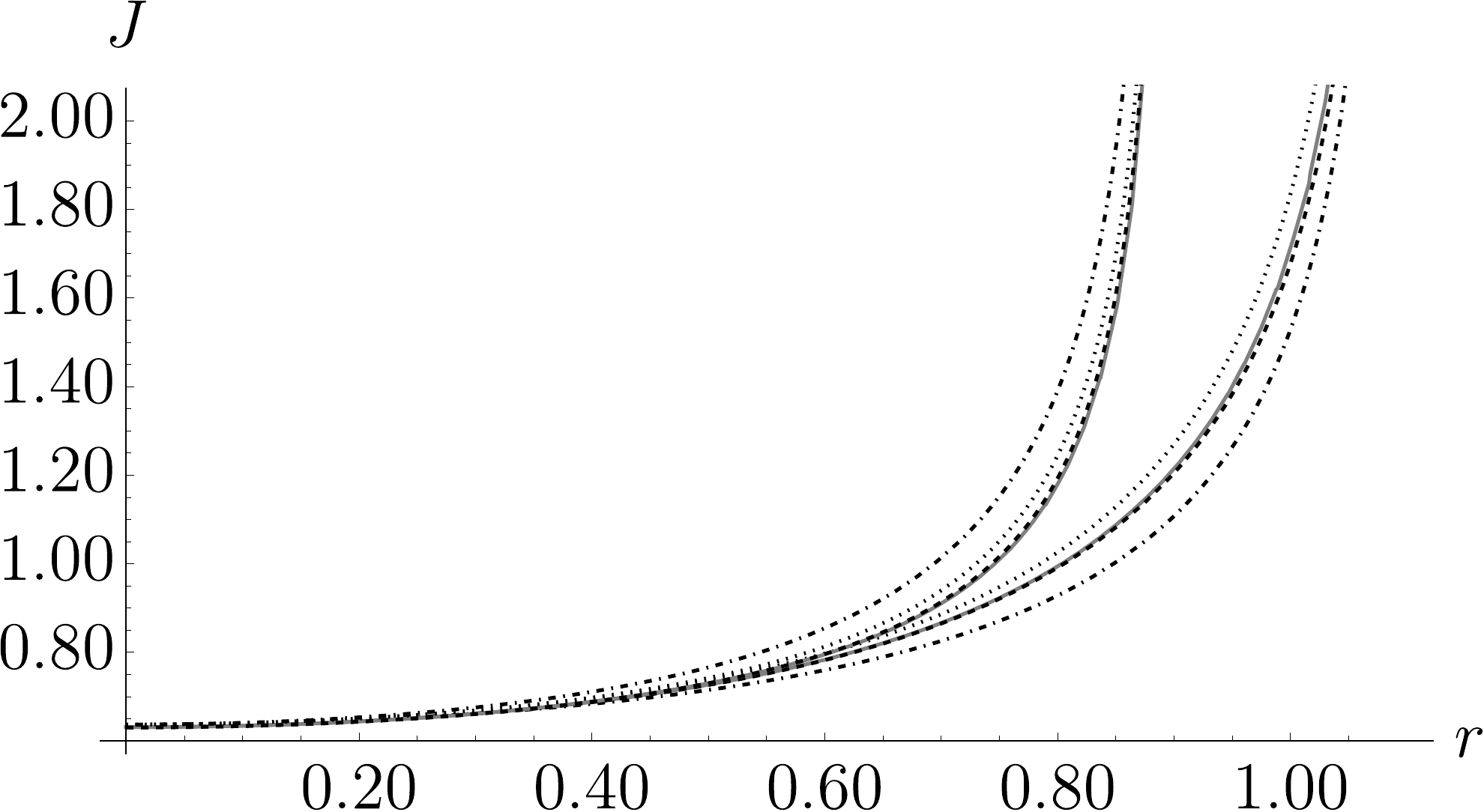}\\
(g) & (h) \\
\end{tabular}
\end{center}
\caption{Validation of \eqref{eq:JSol} with $\epsilon = 0.1$ and, from top-to-bottom, for $n = 3,4,5,6$. (a), (c), (e) and (g) are contour plots of $J$, where the dashed (black) curve is \eqref{eq:JSol} and the solid (gray) curve is according to the results of COMSOL. The solid red curve represents the pinned edge of the droplet. (b), (d), (f) and (h) are plots of $J$ between $r=0$ and the nearest/furthest points on the contact line according to COMSOL (solid curve), leading order (dash-dotted curve), first-order solution (dotted curve), and second order solution (dashed curve) for $J$.}
\label{fig:validation}
\end{figure}

In order to validate these results, we compare against direct numerical simulations (DNS) of the full system (\ref{eq:Laplace})--(\ref{eq:Evap_Flux}) computed using COMSOL Multiphysics (see Appendix \ref{sec:appNumMeth} for further details). Results are displayed for $\epsilon = 0.1$ and different values of $n$ in Figure \ref{fig:validation}.
In Figure \ref{fig:validation} (a), (c), (e), (g), we show contour plots comparing the second-order asymptotic prediction given by (\ref{eq:JSol}) (dashed curves) and the numerical solution (solid curves). It is clear that, even for a relatively large value of $\epsilon$, the agreement is very good, which can be further seen when viewing the evaporative flux along rays of constant $\theta$, as shown in Figure \ref{fig:validation} (b), (d), (f), (h). This agreement holds even close to the contact line, where the flux diverges and we may expect the asymptotic approach to break down due to the caveats surrounding Fabrikant's decomposition.

We can probe the accuracy of (\ref{eq:JSol}) in further detail by considering the error metric 
\begin{equation}
    e(\theta;n)=\frac{1}{0.99+\epsilon \cos(n\theta)}\int_{r=0}^{0.99+\epsilon \cos(n\theta)} \left| \frac{J_\text{asy}-J_\text{num}}{J_\text{num}} \right| \ud r,
    \label{eq:Error}
\end{equation}
where $J_\text{asy}$ and $J_\text{num}$ are the evaporative fluxes according to asymptotic solutions and direct numerical calculations, respectively. Essentially, $e(\theta,n)$ represents an average relative error in the flux prediction along a radial ray. Note that the ray is truncated just inside the contact line --- although past the circle at which the Fabrikant decomposition should break down --- due to the sensitivities associated with the evaporative flux singularity in the COMSOL simulations. Using this metric, we find that, for two specific examples with $\epsilon=0.1$, 
\begin{align}
    e(0;3)&=1.62\times10^{-3},\\
    e(\pi/3;3)&=1.93\times 10^{-3},\\
    e(0;4)&=3.02\times10^{-3},\\
    e(\pi/4;4)&=2.72\times 10^{-3},
\end{align}
where the values of $\theta$ have been chosen since these represent the points furthest and closest to the centre of the droplet, respectively. Clearly, this is further support to the veracity of the asymptotic prediction (\ref{eq:JSol}). Similar levels of agreement persist for other values of $n$.

\section{Application to gravity-dominated droplets}
\label{sec:Large_drops}

Up to this juncture, all of our analysis holds for any thin droplet --- i.e. where the contact radius is much larger than the initial maximum thickness of the droplet. However, to explore the findings of the previous sections in an application, we now turn our focus to large droplets where surface tension is dominated by gravity. Such droplets are often called `pancake' droplets, since the free surface is approximately flat over the bulk of the contact set \cite[][]{rienstra1990shape}, aside from a thin boundary region near the contact line where surface tension is relevant; throughout this analysis we shall assume that surface tension is sufficiently small that this boundary region plays a lower-order role in the analysis so that we may neglect it. We shall periodically revisit the consequences of this assumption in the sequel.

\subsection{Internal flow dynamics}
\label{sec:Flow_Dynamics}

We begin by considering the evaporation-driven flow within the droplet. The analysis of this section holds for pure liquids but also, as we discuss shortly, for droplets containing a solute, provided that it is sufficiently dilute. 

Since the droplet is thin, to leading-order, the droplet free surface $h(r,\theta,t)$, the depth-averaged liquid velocity $\mbf{u}(r,\theta,t) =  u(r,\theta,t)\mathbf{e}_r+v(r,\theta,t)\mathbf{e}_\theta$ and the liquid pressure $p(r,\theta,z,t)$ satisfy the thin-film equations, which, as discussed by, for example, \cite{oliver2015contact}, are given in dimensional form by
\begin{equation}
 \frac{\partial h}{\partial t} + \nabla\cdot\left(h\mbf{u}\right) = -\frac{J}{\rho}, \quad \mbf{u} = -\frac{h^{2}}{3\mu}\nabla p, \quad p = p_{\mathrm{atm}} + \rho g(h-z) - \gamma\nabla^{2}h,
 \label{eq:Dim_Thin_Film}
\end{equation}
where $\nabla = \mbf{e}_{r}\partial/\partial r + \mbf{e}_{\theta}(1/r)\partial/\partial\theta$ is the two-dimensional gradient operator, $\mu$ is the liquid viscosity and $p_{\mathrm{atm}}$ is the ambient pressure of the surrounding gas. Note that the second term in the pressure equation is the hydrostatic pressure, while the final term comes from the curvature of the droplet. 

The system (\ref{eq:Dim_Thin_Film}) is nondimensionalised using the scalings 
\begin{equation}
    r=a_0 \hat{r}, \quad z =\delta a_0 \hat{z}, \quad t=t_\text{ref}\hat{t}, \quad h=\delta a_0 \hat{h},
\end{equation}
\begin{equation}
    \mbf{u}=u_\text{ref}\mbf{\hat{u}}, \quad  w=\delta u_\text{ref} \hat{w}, \quad p-p_a=\rho g a_0 \delta \tilde{p},
\end{equation}
where $t_\text{ref}$ and $u_\text{ref}$ are reference time and velocity scales, given by
\begin{equation}
    t_\text{ref}=\frac{\rho \delta a_0^2}{D(c_\text{sat}-c_\infty)}, \quad u_\text{ref}=\frac{a_0}{t_\text{ref}}=\frac{D(c_\text{sat}-c_\infty)}{\rho \delta a_0}.
\end{equation}
Hence, dropping the caret notation immediately, we find that
\begin{equation}
 \frac{\partial h}{\partial t} + \nabla\cdot\left(h\mbf{u}\right) = -J, \quad \mbf{u} = -\frac{\Bo}{\Ca}\frac{h^{2}}{3}\nabla p, \quad p =  h-z - \frac{1}{\Bo}\nabla^{2}h,
 \label{eq:NonDim_Thin_Film}
\end{equation}
where
\begin{equation}
\Bo = \frac{\rho g a_{0}^{2}}{\gamma} \quad \mbox{and} \quad \Ca = \frac{\mu u_{\mathrm{ref}}}{\delta^{3}\gamma}
\end{equation}
are the Bond and capillary numbers, respectively. 

Assuming the droplet is sufficiently large that $\Bo\gg1$ --- and, for the sake of rigour, that $\Bo\gg \Ca$, although for many typical configurations, the capillary number will be small, \cite[see for example,][for a discussion]{moore2021nascent} --- we seek an expansion of the form
\begin{equation}
    p=h_0 - z + \frac{1}{\Bo} p_0 + O(\Bo^{-2}), \quad  h = h_0 + O(\Bo^{-1}), \quad \mbf{u} = \mbf{u}_0 + O(\Bo^{-1}).
\end{equation}
To leading order, clearly we must have $\nabla (h_0-z) = 0$ so that $h_0=h_0(t)$. Proceeding to the next order, we see that the velocity is given by
\begin{equation}
    \mathbf{u}_0 = -\frac{h_0^{2}}{3\Ca}\nabla p_0,
\end{equation}
where the pressure perturbation $p_0(r,\theta,t)$ is related to the free surface evolution and the evaporative flux via the thin-film equation
\begin{equation}
\dd{h_0}{t}-\frac{h_0^3}{3\Ca}\nabla^2p_0=-J \quad \mbox{for} \quad 0<r<1+\epsilon\cos n\theta, 0\leq \theta<2\pi.
\label{eq:ThinFilm}
\end{equation}
We require a suitable boundary condition to solve this: a physically-reasonable one is to impose no-flux of liquid through the pinned contact line,
\begin{equation}
 -\mathbf{n}\cdot\frac{h_0^3}{3\Ca}\nabla p_0 = 0 \quad \mbox{on} \quad r = 1+\epsilon\cos n\theta, 0\leq \theta<2\pi
 \label{eq:NoFlux}
\end{equation}
where $\mathbf{n}$ is the outward-pointing unit normal to the contact line. 

Hence, given the evaporative flux (\ref{eq:JSol}), the leading order flow dynamics reduces to solving (\ref{eq:ThinFilm})--(\ref{eq:NoFlux}), which we shall now pursue. In so doing, we drop the subscript notation on the leading order variables for brevity.

\subsubsection{Free surface profile and droplet lifetime}

To determine $h(t)$, we consider the usual liquid mass conservation equation $dV/dt=-F$, which yields
\begin{equation}
\dd{}{t}\left[ h(t)\int_{\theta=0}^{2\pi}\int_{r=0}^{1+\epsilon \cos(n\theta)}r\ud r\ud \theta \right]=-\int_{0}^{2\pi}\int_{0}^{1+\epsilon\cos n\theta} rJ(r,\theta)\ud r\ud \theta = -(4+\epsilon^{2}n).
\label{eq:LiqMassConv1}
\end{equation}
by (\ref{eq:F}). Hence, evaluating the integral on the left-hand side of (\ref{eq:LiqMassConv1}), we find
\begin{equation}
    \pi\left(1+\frac{\epsilon^2}{2}\right)\dd{h}{t}+\od{3}=-(4+\epsilon^2n)+\od{3} \quad \implies \quad h=1-\frac{t}{\pi}\left(4+\epsilon^2(n-2)\right)+\od{3}.
\end{equation}
We therefore find that the droplet lifetime --- that is, the time at which all the liquid has fully evaporated --- is given up to second order in $\epsilon$ by
\begin{equation}
    t=t_f=\frac{\pi}{4}\left(1-\frac{\epsilon^2}{4}(n-2)\right).
\end{equation}
Notably, as for the expansion of the evaporative flux in the vicinity of the internal stagnation point of the droplet (\ref{eq:JStagPoint}), the perturbation to the free surface evolution and the droplet lifetime are significantly smaller for the $n = 2$ case than for $n>2$. We also note that the conservation condition (\ref{eq:LiqMassConv1}) also guarantees the existence of a solution to the Neumann problem (\ref{eq:ThinFilm})--(\ref{eq:NoFlux}).

\subsection{Liquid pressure}
\label{sec:Pressure} 

The liquid pressure may now be calculated from (\ref{eq:ThinFilm})--(\ref{eq:NoFlux}). To simplify the algebra, we write
\begin{equation}
    p=\frac{3\Ca}{h^3}\left[ \frac{\dot{h}}{4}r^2+\frac{2}{\pi}Q \right] \qquad \implies\qquad  \nabla^2Q=\frac{\pi}{2}J. 
    \label{eq:PressureFull}\\
\end{equation}

\subsubsection{Outer region}

In the bulk of the droplet where $1-r = O(1)$, the expression (\ref{eq:JSol}) for evaporative flux may be expanded as $J = J_{0}(r) + \epsilon J_{1}(r,\theta) + \epsilon^{2} J_{2}(r,\theta) + \dots$ as $\epsilon\rightarrow0$, where
\begin{alignat}{2}
J_{0}(r) & \; = && \; \frac{2}{\pi\sqrt{1-r^{2}}}\\
J_{1}(r,\theta) & \; = && \; \frac{2}{\pi(1-r^{2})^{3/2}}\left((1-r^{2})f_{1}(r;n) - r^{2}\right)\cos{n\theta}\\
J_{2}(r,\theta) & \; = && \frac{1}{\pi(1-r^{2})^{5/2}}\left[2(1-r^{2})^{2}(f_{20}(r;n) + f_{22}(r;n)\cos{2n\theta} + \frac{r^{2}}{2}(1+\cos{2n\theta})(3-2f_{1}(r;n)(1-r^{2}))\right] \; 
\end{alignat}
where $f_{1}$, $f_{20}$ and $f_{22}$ are given by (\ref{eq:f1})--(\ref{eq:f22}). This suggests seeking a solution for $Q$ of the form
\begin{alignat}{2}
    Q & = && Q_0(r)+\epsilon (C_{O10} + Q_1(r)\cos n\theta)+\epsilon^2(Q_{20}(r)+Q_{21}(r)\cos 2n\theta) + \dots
\end{alignat}
as $\epsilon\rightarrow0$. We find that
\begin{alignat}{2}
    Q_0(r) &\; = && \; C_{O0}-\sqrt{1-r^2}+\tanh ^{-1}\left(\sqrt{1-r^2}\right)+\log (r),\label{eq:QOuter} \\
    Q_1(r;n) &\; = && \; C_{O11} r^n+\frac{1}{4} r^{n+2} \Gamma (n) \, _2\tilde{F}_1\left(\frac{3}{2},n+1;n+2;r^2\right)-\frac{r^n}{2 n \sqrt{1-r^2}}, \label{eq:Q1Outer}\\
    Q_{20}(r;n) &\; = && \; C_{O20}+\frac{1}{4} \left(n \log \left(\sqrt{1-r^2}+1\right)+\frac{1}{\sqrt{1-r^2}}\right),\label{eq:Q20Outer}\\
    Q_{21}(r;n) &\; = && \; C_{O21} r^{2 n}-\frac{r^{2 n} \left((2 n-1) r^2-2 n\right)}{16 n \left(1-r^2\right)^{3/2}}+\nonumber \\
    &\;  && \; \frac{r^{2 n} \left(r^2\right)^{-2 n}
   \left((2 n-1) B_{r^2}\left(2 n+2,-\frac{3}{2}\right)-2 (n+1) B_{r^2}\left(2 n+1,-\frac{3}{2}\right)\right)}{32 n},\label{eq:Q21Outer}
\end{alignat}
where $B_{x}(\alpha,\beta) = \int_{0}^{x}t^{\alpha-1}(1-t)^{\beta-1}\,\mbox{d}t$ is the incomplete beta function and $_2\tilde{F}_1\left(a,b;c;d\right) = {}_2F_1\left(a,b;c;d\right)/\Gamma(c)$ is the regularised hypergeometric function. The constants $C_{O0}$, $C_{O10}$ and $C_{O20}$ are arbitrary undetermined functions of time (recall that the pressure is determined by solving a Neumann problem). On the other hand, the coefficients $C_{O11}$ and $C_{O21}$ must be determined by matching to an inner region in the vicinity of the contact line where our na\"{i}ve expansions for $J$ and $Q$ break down.  

\subsubsection{Inner region}\label{sec:innerRegionFluxAndPressure}

We introduce the local variable
\begin{equation}
    r=1-\epsilon \bar{r}. \label{eq:Inner_Scale}
\end{equation}
Upon substituting this into the evaporative flux (\ref{eq:JSol}) and expanding as $\epsilon\rightarrow0$, we find that the local expansion of the evaporative flux is given by
\begin{equation}
    J=\frac{1}{\sqrt{\epsilon}}\left(J_0(r,\theta)+\epsilon J_1(r,\theta)+\epsilon^2J_2(r,\theta)+\cdots\right),
    \label{eq:J_inner_exp}
\end{equation}
where
\begin{align}
    J_0(r,\theta)&=\frac{\sqrt{2}}{\pi  \sqrt{\cos (n \theta )+\br}},\\
    J_1(r,\theta)&=\frac{(2 n-1) \cos (n \theta )+\br}{2 \sqrt{2} \pi  \sqrt{\cos (n \theta )+\br}}, \\
    J_2(r,\theta)&=\frac{-4 (2 n (2 n-5)+3) \br \cos (n \theta )+(3-4 n (3 n+1)) \cos (2 n \theta )-4 (n-1) n+6 \br^2+3}{32 \sqrt{2} \pi  \sqrt{\cos (n \theta
   )+\br}}.
\end{align}
Upon substituting the scaling (\ref{eq:Inner_Scale}) into the Poisson equation (\ref{eq:ThinFilm}), we find that 
\begin{equation}
     \left(\frac{1}{\epsilon^2}\q{^2}{\bar{r}^2}-\frac{1}{1-\epsilon \bar{r}}\q{}{\bar{r}}+\frac{1}{(1-\epsilon \bar{r})^2}\q{^2}{\theta^2}\right) Q^{(i)}=\frac{\pi}{2}J,
\end{equation}
for $\br > -\cos{n\theta}, \; 0\leq\theta<2\pi$, while the no-flux condition (\ref{eq:NoFlux}) is given by
\begin{equation}
-1-\epsilon \cos n\theta-\frac{1}{\epsilon}\q{Q^{(i)}}{\bar{r}}+\epsilon\q{^2Q^{(i)}}{\theta^2}+\epsilon^2\frac{2-n}{4}=0 \quad \mbox{on} \quad \br = -\cos{n\theta}, \; 0\leq\theta<2\pi;
\end{equation}
together these suggest that we seek an inner expansion of the form
\begin{equation}
    Q^{(i)}=C_{O0}+\epsilon\left( Q^{(i)}_0(\br,\theta)+\epsilon^{1/2}Q^{(i)}_{1/2}(\br,\theta)+\epsilon Q^{(i)}_1(\br,\theta)+\cdots\right)
    \label{eq:Q_inner_exp}
\end{equation}
as $\epsilon\rightarrow0$. Proceeding order-by-order in the standard way, we find the first five inner solutions are given by
\begin{align}
    Q^{(i)}_0&=C_{I0}(\theta)-\bar{r}, \label{eq:Q0Inner}\\
    Q^{(i)}_{1/2}&=C_{I1/2}(\theta)+\frac{2\sqrt{2}}{3}  (\cos (n \theta )+\bar{r})^{3/2}, \label{eq:Q1p5Inner}\\
    Q^{(i)}_1&=C_{I1}(\theta)-2 \bar{r} \cos (n \theta )-\frac{\bar{r}^2}{2}, \label{eq:Q1Inner}\\
    Q^{(i)}_{3/2}&=C_{I3/2}(\theta)+\frac{((10 n-1) \cos (n \theta )+9 \bar{r}) (\cos (n \theta )+\bar{r})^{3/2}}{15 \sqrt{2}},\\
    Q^{(i)}_2&=C_{I2}(\theta)-\frac{1}{6} \bar{r} \left(6 \cos (n \theta ) \left(C_{I0}''(\theta )+\bar{r}\right)-6 n \sin
   (n \theta ) C_{I0}'(\theta )+3 \cos (2 n \theta )+3 \bar{r} C_{I0}''(\theta )+2 \bar{r}^2+3\right).
\end{align}

It remains to determine the constants $C_{O11}$, $C_{O21}$ from the outer region \eqref{eq:QOuter} and the functions $C_{Ij}(\theta)$ from the inner. This may be done by using a standard matching procedure (using, for example, an intermediate variable), and we find that
\begin{align}
    C_{O11}&=\frac{2}{n}-\frac{\sqrt{\pi } \Gamma (n)}{ 2\Gamma \left(n+\frac{1}{2}\right)},\\
    C_{O21}&=\frac{1}{8} \left(\frac{2}{n}+\sqrt{\pi } \left(\frac{4 \Gamma (n+1)}{\Gamma \left(n+\frac{1}{2}\right)}-\frac{\Gamma (2 n+1)}{\Gamma
   \left(2 n+\frac{1}{2}\right)}\right)-8\right),\\
   C_{I1/2}(\theta)&=C_{I3/2}(\theta)=0,\\
    C_{I0}(\theta)&= C_{O10} + \left(\frac{2}{n}-\frac{\sqrt{\pi } \Gamma (n)}{\Gamma \left(n+\frac{1}{2}\right)}\right) \cos (n \theta),\\
    C_{I1}(\theta)&=C_{O20} + \frac{\cos(2n\theta)}{4} \left[\frac{1}{n}-4+\sqrt{\pi } \left(\frac{2 n \Gamma (n)}{\Gamma
   \left(n+\frac{1}{2}\right)}-\frac{\Gamma (2 n+1)}{\Gamma \left(2 n+\frac{1}{2}\right)}\right) \right].
\end{align}
The remaining unknown function $C_{I2}(\theta)$ may be determined by proceeding to higher order in the outer region, but since it is not needed in the present analysis, we shall forgo this.

\subsection{Final residue}
\label{sec:Residue}

Having determined the flow dynamics, we may now investigate the transport of an inert, dilute solute within the droplet whose concentration is given by $\phi = \phi_{\mathrm{ref}}\hat{\phi}$, where $\phi_{\mathrm{ref}}$ is the initial (uniform) solute concentration. We shall, once again, drop the caret on the dimensionless variable going forward.

Under the dilute assumption, the flow and solute transport completely decouple and the distribution of the solute may be analysed in detail by considering an appropriate advection-diffusion equation for the solute concentration within the droplet \cite[see, for example,][]{wray2014electrostatic,saenz2017dynamics,moore2021nascent,moore2022nascent}. However, in the present analysis, we shall focus on the final residue at the contact line --- the `coffee ring' --- which simplifies the analysis. 

Previous studies \cite[][]{FreedBrown2015,saenz2017dynamics,moore2022nascent} have demonstrated that asymmetry in the contact line geometry leads to heterogeneities in the coffee ring profile. This holds for both surface-tension dominated and gravity-dominated droplets. In particular, the coffee ring effect is enhanced (respectively, inhibited) in regions where the curvature of the contact line is larger (smaller). This effect can be shown to arise purely due to the geometry of the droplet by considering a uniform evaporative flux \cite[][]{FreedBrown2015,moore2022nascent}, but is further exacerbated in regimes where evaporation is diffusion-dominated, due to the enhanced flux in these high-curvature regions \cite[see, for example,][but also note the plots in Figure \ref{fig:validation}]{saenz2017dynamics}. 

Here, we investigate this phenomenon quantitatively for a range of different droplet geometries. As stated above, we shall concentrate on the variation of the final residue at the contact line as a function of position. In our analysis, we shall neglect the effects of any solute jamming or coupling between the flow and transport problems --- this is very likely to be important at later stages of the evaporative process for a real world scenario, see, for instance, \cite{popov2005evaporative,kaplan2015evaporation,Guazzelli2018} for a discussion of different models for finite particle size effects, and \cite{moore2021nascent,moore2022nascent} for a discussion of the limits of the dilute regime --- so that, by the time the droplet fully evaporates, all of the solute has been transported to the contact line \cite{deegan1997capillary}. We may exploit this fact to greatly simplify our calculations. 

\begin{figure}
\begin{center}
\begin{tabular}{c}
\includegraphics[width=0.4\textwidth]{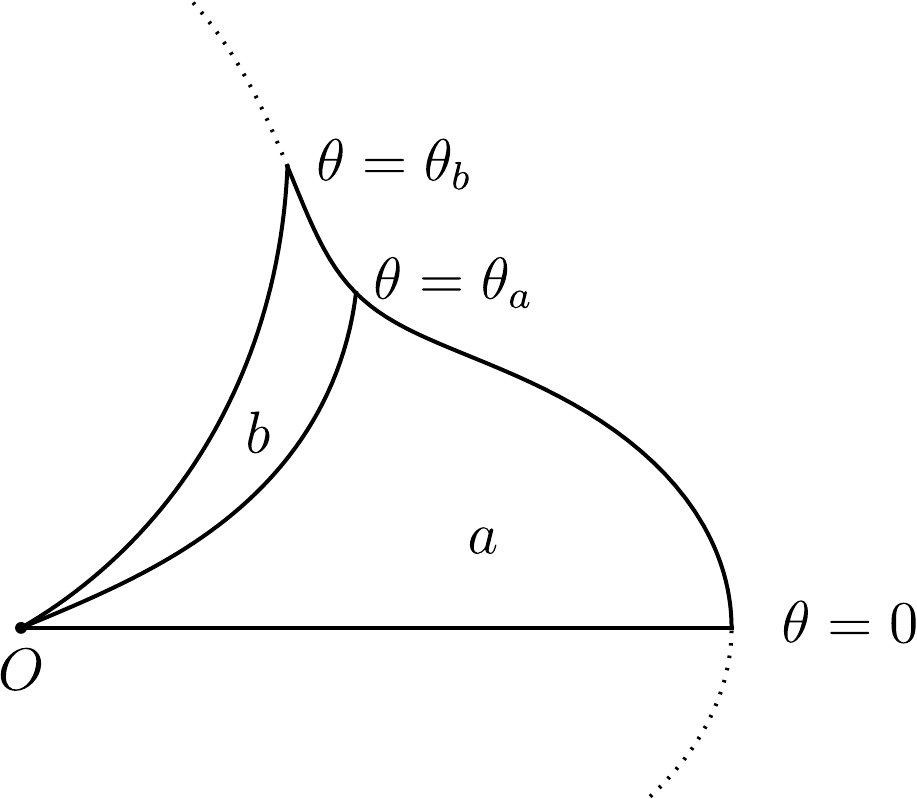}
\end{tabular}
\end{center}
\caption{Interfacial configuration for $n=4$.}
\label{fig:regionPlot}
\end{figure}

In the thin-droplet limit, vertical diffusion of solute is sufficiently strong that the distribution of solute is independent of $z$ to leading order and, as shown by, for example, \cite{wray2021contact}, the solute concentration is advected along particle paths, viz:
\begin{equation}
    \dd{\phi}{t}=\frac{\phi  J}{h} \quad \text{on} \quad \dd{r}{t}=u, \quad \dd{\phi}{t}=\frac{v}{r}.
\end{equation}
Moreover, since pathlines coincide with streamlines due to the separable nature of time in the flow, we can actually simplify things even further. Consider the situation in Figure \ref{fig:regionPlot}, where two streamlines $\theta = \psi(r;\theta_i)$ start at the source at the centre of the droplet and, by dryout, reach the contact line at $\theta=\theta_i$ where $i = a, b$, respectively. Then, the total  mass accumulated at the contact line between $\theta_a$ and $\theta_b$ is exactly the total initial mass in region $b$. As a result, the density of the final mass accumulated at the contact line, $D$, is given by
\begin{align}
D&=\dd{M}{s}=\left.\frac{1}{\mathrm{d}s/\mathrm{d}\theta}\dd{M}{\theta}\right|_{\theta = \theta_a}=\left.\frac{1}{\sqrt{r^2+r_\theta^2}}\dd{M}{\theta}\right|_{\theta = \theta_a}=\left.\frac{1}{\sqrt{a^2+a_{\theta}^2}}\dd{M}{\theta}\right|_{\theta = \theta_a}.
\label{eq:Density_Full}
\end{align}

Thus, we have reduced the problem to determining $\ud M/\ud\theta$ at the contact line. Let $\ar(\theta_a)$ be the area bounded by $\theta=0$, the contact line, and the streamline $\theta=\psi(r;\theta_a)$, so that in Figure \ref{fig:regionPlot}, $\ar(\theta_a)$ corresponds to region $a$. Then,
\begin{align}
\left.\dd{M}{\theta_a}\right|_{\theta_a}&=\lim_{\delta \theta\to0} \left[\iint_{\ar(\theta_a+\delta\theta)}\ud A-\iint_{\ar(\theta_a)}\ud A\right]=\lim_{\delta \theta\to0}\int_{r=0}^{1+\epsilon \cos(n\theta_a)} \int_{\theta=\psi(r;\theta_a)}^{\psi(r;\theta_a+\delta \theta)}r\ud r\ud \theta\\
&=\int_{r=0}^{1+\epsilon \cos(n\theta_a)} r\q{\psi}{\theta_a}\ud r.
\end{align}
Thus, we now need to write $\psi$ as a function of $\theta_a$ (and $r$), which we pursue using the differential equation
\begin{equation}
\dd{\psi}{r}=\frac{v}{ru}=\frac{p_\theta}{r^2p_r} \quad \mbox{for} \quad 0<r<1+\epsilon\cos{n\theta_a}\label{eq:psiEqn}
\end{equation}
subject to $\psi(1+\epsilon\cos{n\theta_a};\theta_a) = \theta_a$. Given that the pressure solution found in \textsection \ref{sec:Pressure} is given in two distinct regions, we find a similar asymptotic structure holds for the streamlines.

\subsubsection{Outer region}

Now, since, for a nearly-circular droplet the streamlines will be small perturbations to a radial ray, we seek an asymptotic solution of the form
\begin{equation}
\psi=\theta_a+\epsilon \psi_1+\epsilon^2\psi_2+\cdots,
\end{equation}
so that \eqref{eq:psiEqn} gives
\begin{equation}
    \dd{\left(\epsilon\psi_1+\epsilon^2\psi_2\right)}{r}=
    \epsilon\frac{ n Q_1 \sin \left(n \theta_a\right)}{r^2
   \left(r-Q_0'\right)}
   +\epsilon^2 \left[\frac{n^2  Q_1 \cos \left(n \theta_a\right)}{r^2 \left(r-Q_0'\right)}\psi_1+\frac{2 n Q_{21} \sin
   \left(2 n \theta_a\right)}{r^2 \left(r-Q_0'\right)}+\frac{n Q_1 Q_1' \sin \left(2n \theta_a\right) }{2r^2 \left(r-Q_0'\right)^2}\right],
\end{equation}
where $Q_{0}$, $Q_{1}$ and $Q_{21}$ are given by (\ref{eq:QOuter}), (\ref{eq:Q1Outer}) and (\ref{eq:Q21Outer}), respectively.

Unfortunately, analytic solutions for $\psi_1$ and $\psi_2$ are not available for general $n$ --- although for a specific $n$, asymptotic solutions and numerical solutions are relatively straightforward to find. Here, for illustrative purposes, we examine the particular case $n=2$. We find
\begin{equation}
\psi_1=\frac{\sin 2\theta_a}{3}\left[P_{O1} + \frac{1}{r^{2}} \left(\sqrt{1-r^2}+r^2 \left(\tanh ^{-1}\left(\sqrt{1-r^2}\right)+2 \log
   (r)\right)-1\right)\right],
   \label{eq:outerPsi2n1}
\end{equation}
where $P_{O1}$ must be determined by matching. Similarly, at $\od{2}$, we find
\begin{align}
    \psi_2=&\frac{\sin 4 \theta_a}{630} \left[P_{O2}-\frac{4}{r^4}+\frac{3 r^2}{2}-\frac{41+70P_{O1}}{r^2}+35 \log ^2\left(\sqrt{1-r^2}+1\right)-5(14P_{O1}-33)\log
   \left(\sqrt{1-r^2}+1\right)\right.\\
   &+\left.\frac{70 \left(\sqrt{1-r^2}-1\right) \log \left(\sqrt{1-r^2}+1\right)}{r^2}+\frac{70 \log (r)
   \left(\sqrt{1-r^2}+r^2 \log \left(\sqrt{1-r^2}+1\right)-1\right)}{r^2}\right.\\
   &\left.+\sqrt{1-r^2}
   \left(\frac{4}{r^4}+\frac{43+70P_{O1}}{r^2}+\frac{210}{1-r^2}-273\right)+35 \log ^2(r)+35(1+2P_{01})\log (r)\right], \label{eq:outerPsi2n2}
\end{align}
where $P_{02}$ again must be determined by matching.

As expected, this asymptotic solution becomes disordered close to the contact line, so we turn to a boundary layer to determine the local streamlines and find the constants $P_{O1}$ and $P_{O2}$.

\subsubsection{Inner region}

Again utilising the scaling (\ref{eq:Inner_Scale}) and the expansion (\ref{eq:Q_inner_exp}) for $Q$, we seek an asymptotic solution of (\ref{eq:psiEqn}) of the form
\begin{equation}
\psi=\theta_a+\epsilon^{3/2}\left(\psi^{(i)}_0+\epsilon^{1/2}\psi^{(i)}_{1/2}+\cdots\right)
\end{equation}
as $\epsilon\rightarrow0$. The streamline equation (\ref{eq:psiEqn}) becomes
\begin{equation}
    -\frac{1}{\epsilon}\dd{\left(\epsilon^{3/2}\psi^{(i)}_0+\epsilon^2\psi^{(i)}_{1/2}\right)}{\bar{r}}=\epsilon^{1/2}\left[ -\frac{\partial Q^{(i)}_0/\partial \theta}{\partial Q^{(i)}_{1/2}/\partial \bar{r}} \right]
    +\epsilon\left[
        -\frac{\partial Q^{(i)}_{1/2}/\partial \theta}{\partial Q^{(i)}_{1/2}/\partial \bar{r}}+\q{Q^{(i)}_0}{\theta}\left(\q{Q^{(i)}_{1/2}}{\bar{r}} \right)^{-2}\left( \q{Q^{(i)}_1}{\bar{r}}-\bar{r} \right)
    \right],\nonumber
\end{equation}
for $\br>-\cos n\theta_a$, where $Q^{(i)}_0$, $Q^{(i)}_{1/2}$ and $Q^{(i)}_1$ are given by (\ref{eq:Q0Inner}), (\ref{eq:Q1p5Inner}) and (\ref{eq:Q1Inner}), respectively. This must be solved subject to the boundary conditions
\begin{equation}
 \psi_{0}^{(i)}(-\cos{n\theta_a}) = \psi_{1/2}^{(i)}(-\cos{n\theta_a}) = 0.
\end{equation}

The inner problem is tractable for general $n$: solving successively yields
\begin{align}
    \psi^{(i)}_0&=\sqrt{2}\frac{ \left(\sqrt{\pi } \Gamma (n+1)-2 \Gamma \left(n+\frac{1}{2}\right)\right) }{\Gamma \left(n+\frac{1}{2}\right)}\sqrt{\cos
   \left(n \theta _a\right)+\bar{r}}\,\sin n \theta _a ,\\
   \psi^{(i)}_{1/2}&=\left(-n+\frac{\sqrt{\pi } \Gamma (n+1)}{\Gamma \left(n+\frac{1}{2}\right)}-2\right)  (\cos (n \theta_a )+\bar{r})\sin n \theta_a.
\end{align}
In the particular case $n=2$, the first two inner solutions are thus given by
\begin{align}
    \psi^{(i)}_0&=\frac{2}{3} \sqrt{2} \sqrt{\cos \left(2 \theta _a\right)+\bar{r}}\, \sin 2 \theta _c , \label{eq:Inner_Psi_1}\\
   \psi^{(i)}_{1/2}&=-\frac{4}{3}  (\bar{r}+\cos (2 \theta_a ))\sin 2 \theta_a.
   \label{eq:Inner_Psi_2}
\end{align}
These can be matched against \eqref{eq:outerPsi2n1}--(\ref{eq:outerPsi2n2}) in a similar manner to the pressure, yielding
\begin{equation}
    P_{O1} = 1, \quad P_{O2}=-\frac{613}{2}.
\end{equation}

Hence, for $n = 2$, we may construct a additive composite solution for the streamlines by combining (\ref{eq:outerPsi2n1})--(\ref{eq:outerPsi2n2}) and (\ref{eq:Inner_Psi_1})--(\ref{eq:Inner_Psi_2}), which takes the form
\begin{alignat}{2}
    \psi & \; = && \;\theta_a+\epsilon \psi_1+\epsilon^{2} \psi_2+\epsilon^{3/2}\psi^{(i)}_0\left( \frac{1-r}{\epsilon} \right)+\epsilon^2\psi^{(i)}_{1/2}\left( \frac{1-r}{\epsilon} \right)\nonumber \\
    & \; && \; -\left[ \epsilon\frac{2\sqrt{2}}{3} \sqrt{1-r}\, \sin 2\theta _a-\epsilon\frac{4}{3} (1-r) \sin2 \theta _a+\epsilon^2\frac{\sin 4 \theta _a}{3\sqrt{2} \sqrt{1-r}} + \frac{4}{3}\cos4\theta_a\right],
    \label{eq:CompositeStreamlines}
\end{alignat}
where the terms in the second line represent the overlap contributions between the outer and inner solution, which have been determined using Van Dyke's matching rule \cite[][]{VanDyke1964}.

\subsubsection{Mass determination}

Finally, we wish to compute
\begin{equation}
    \left.\dd{M}{\theta}\right|_{\theta=\theta_a}=\int_0^{1+\epsilon \cos n \theta_a}r\q{\psi}{\theta_a}\ud r
    \label{eq:Integral_Dummy}
\end{equation}
and hence the density $D$ given by \eqref{eq:Density_Full}. Since we only have a complete analytical solution for $n = 2$, we shall present the residue calculation in detail for this example. It is straightforward to extend the analysis to a particular $n$, although we are unable to present a closed form solution for general $n$.

Given that we have inner and outer solutions for $\psi$, we can divide (\ref{eq:Integral_Dummy}) into two integrals by choosing $0<\epsilon\ll\delta\ll1$, and then splitting the interval of integration into $(0,1-\delta)$ and $(1-\delta,1+\epsilon \cos 2\theta_a)$, where we use the outer solution for $\psi$ in the first interval, and the inner solution in the second. Thus, the density is given by
\begin{alignat}{2}
    D & \; = && \; \frac{1}{\sqrt{\left(1+\epsilon \cos 2\theta_a\right)^2+\left(-\epsilon2\sin 2\theta_a\right)^2}}\int_0^{1+\epsilon \cos n \theta_a}r\q{\psi}{\theta_a}\ud r \\
    & \; \sim && \;\left(1 - \epsilon\cos2\theta_a + \frac{\epsilon^2}{2}(\cos4\theta_a -1)\right)\left[\int_0^{1-\delta} r\q{}{\theta_a}\left[\theta_a+\epsilon \psi_1+\epsilon^2 \psi_2 \right]\ud r +\right.\\
    & \; && \; \left.\int_{\delta/\epsilon}^{-\cos 2 \theta_a}(1-\epsilon \bar{r})\q{}{\theta_a}\left[ \theta_c+\epsilon^{3/2}\psi^{(i)}_0+\epsilon^2\psi^{(i)}_2 \right]\left(-\epsilon \ud \bar{r}\right)\right]\\
    & \; \sim && \; \frac{1}{2}+ \frac{\epsilon}{6} (1+4\log 2) \cos 2 \theta _a-\epsilon^2\left[ \frac{1+\log 2}{3}+\frac{501+70 \pi ^2+1116 \log 2-1680 \log ^2 2}{3780}\cos 4 \theta_a \right]. \label{eq:Density_Sol}
\end{alignat}
as $\epsilon\rightarrow0$.

There are several interesting points about the preceding result. Firstly, if $\epsilon=0$, we retrieve the expected uniform density $D = 1/2$ for a circular droplet. Secondly, when $\epsilon>0$, we see that the coefficient of the $O(\epsilon)$-term in (\ref{eq:Density_Sol}) is maximized for $\theta_a = 0, \pi$ --- i.e. where the contact line has highest curvature --- and minimized when $\theta_a = \pi/2, 3\pi/2$ --- i.e. where the contact line has lowest curvature. Thus, we see the expected enhancement (respectively, diminishing) of the coffee ring effect along the high-curvature (low curvature) parts of the contact line. 

Thirdly, we note that, in the vicinity of zeroes of $\cos2\theta_a$, the $O(\epsilon^2)$-term dominates the $O(\epsilon)$-term in (\ref{eq:Density_Sol}). Indeed, we moreover find that there are four regions in which the perturbation of the density from the circular solution is $o(\epsilon^2)$; namely about
\begin{equation}
\theta_a = \frac{\pi}{4} + \epsilon\alpha, \; \frac{3\pi}{4} - \epsilon\alpha, \; \frac{5\pi}{4} + \epsilon\alpha, \; \frac{7\pi}{4} - \epsilon\alpha \quad \mbox{where} \quad \alpha = \frac{70\pi^2 - 1680\log^2{2} - 144\log{2} - 759}{1260(4\log{2} + 1)}. 
\end{equation}

To calculate the cumulative mass, $M(\theta_a)$, between the ray $\theta = \theta_a$ and the horizontal, we may integrate the density (\ref{eq:Density_Sol}) around the contact line with respect to arc length $s$. We find that 
\begin{equation}
    M=\theta_a\left(\frac{1}{2}+\epsilon^2 c_1\right)+\epsilon c_2 \sin 2\theta_a+\epsilon^2 c_3\sin 4\theta_a,\label{eq:gen2Mass}
\end{equation}
where the coefficients $c_1$, $c_2$ and $c_3$ are given in Table \ref{tab:secondOrderComparison}. For reference, we also include numerical estimates of the same coefficients for $\epsilon = 0.2$. We see that, despite the relatively large value of $\epsilon$, the comparison between the numerical and asymptotic predictions of the final residue is good. 

\begin{figure}
\begin{center}
\begin{tabular}{c}
\includegraphics[width=0.6\textwidth]{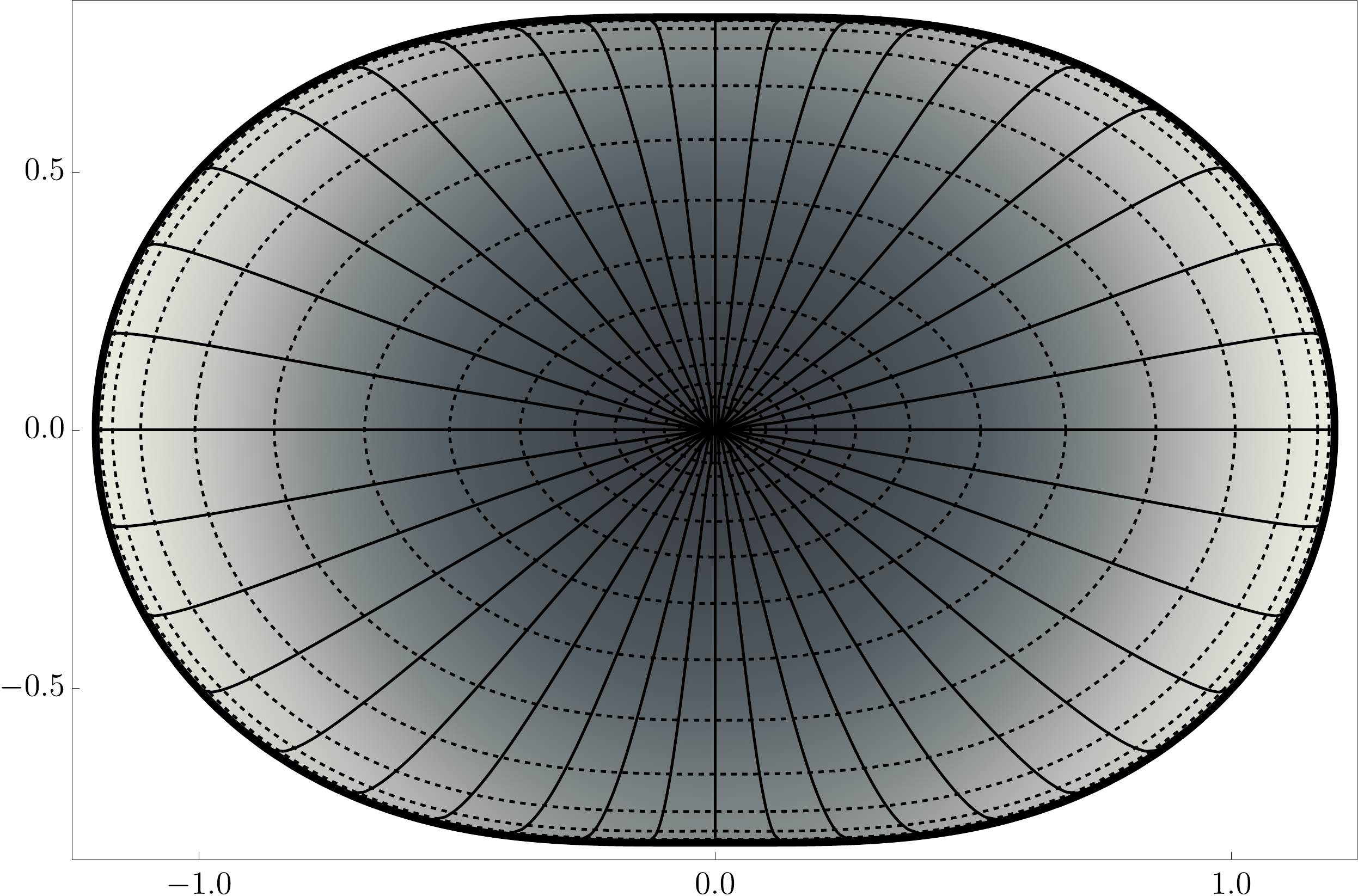}
\end{tabular}
\end{center}
\caption{Asymptotic for the liquid pressure (darker shading corresponds to higher pressure), contours of flux (dashed curves) and liquid streamlines (solid curves) for the case $n=2$, $\epsilon=0.2$.}
\label{fig:n2Plot}
\end{figure}

Further details of the asymptotic solution for the $n = 2$ example are shown in Figure \ref{fig:n2Plot}. We plot contours of the evaporative flux (dashed lines), while the liquid pressure (\ref{eq:PressureFull}) is given by the shading, with the higher pressure corresponding to darker shading. The pressure gradient drives a volume flux towards the contact lines, with the composite streamlines (\ref{eq:CompositeStreamlines}) given by  the solid curves. Note that the streamlines do not approach the contact line perpendicularly as observed in the surface tension-dominated limit \citep{saenz2017dynamics,wray2021contact}. This is due the absence of the $h\sim a-r$ zero of the interface in the present work due to the neglect of the small surface tension boundary layer. In practice, we anticipate that including this boundary layer would result in such behaviour being recovered.

\begin{table}
\centering
\def~{\hphantom{0}}
\begin{tabular}{ c  c  c  c }
 Method & $c_1$ & $c_2$ & $c_3$\\ 
 \hline 
 Asymptotics & $\frac{1}{4}=0.25$ & $\frac{1}{3}\left(1+\log 2\right)\approx0.5644$ & $\frac{-1038-35\pi^2+72\log 2+840\log^2 2}{7560}\approx -0.1230$  \\ 
Numerics using $\epsilon=0.2$ & 0.2478 & 0.5646 & -0.1152
\end{tabular}
\caption{\label{tab:secondOrderComparison}Coefficients for the cumulative final masss at the contact line as a function of angle \eqref{eq:gen2Mass} according to the asymptotics and numerics.}
\end{table}

\section{Droplets with polychromatic footprints}

\subsection{Evaporation of droplets with regular polygonal footprints}
\label{sec:Polygon_Fluxes}

As discussed previously, it is very desirable to be able to determine the evaporative flux for droplets with non-monochromatic footprints, especially shapes such as polygons. However, exact polygons are problematic due to the presence of sharp corners (and hence Fourier representations that do not decay), and so instead we apply our approach of \textsection\textsection \ref{sec:Evap_Flux}--\ref{sec:Large_drops} to smoothed polygons. We note that \cite{popov2003characteristic} and \cite{zheng2005deposit} discuss the evaporative flux and associated deposition pattern for solute-laden flows in sharp corners in detail. 

To utilize our analysis in \textsection \ref{sec:Evap_Flux}, we seek a Fourier representation of a smoothed polygon as follows. The parametric equation for a regular $n$-gon is 
\begin{equation}
    r=\left| \sec \left( \theta-\frac{2\pi}{n}\left\lfloor \frac{\pi+n\theta}{2\pi} \right\rfloor \right) \right|.
\end{equation}
This is evolved under the heat equation
\begin{equation}
    \q{f_s^{(i)}}{t}=\frac{1}{r^2}\q{^2f_s^{(i)}}{\theta^2},
\end{equation}
while being normalised to have constant term unity, until the maximum curvature of $f_s$ does not exceed 10 dimensionless units. This is then taken to be the shape of the droplet $a(\theta)$. An expression for the evaporative flux is then sought in the form
\begin{equation}
    J=\frac{2}{\pi}\frac{a}{\sqrt{a^2-r^2}}\left\{1+\sum_{i=1}^{N_1}c_{ni} f_1(r;ni)\cos ni\theta+\sum_{i=1}^{N_2}c_{ni}^2f_{20}(r;ni)+\sum_{i=1}^{N_3}c_{ni}^2f_{22}(r;ni)\cos 2ni\theta \right\},
    \label{eq:JSol_Gen}
\end{equation}
where $f_{1}$, $f_{21}$ and $f_{22}$ are given by (\ref{eq:f1}), (\ref{eq:f20}) and (\ref{eq:f22}), respectively and the $N_1$, $N_2$ and $N_3$ are selected to minimise $e(0,n)+e(\pi/n,n)$. In all cases we find $N_2=N_3$, and the corresponding $N_1$ and $N_2$ for different polygons are given in Table \ref{tab:smoothing}. Note that for higher polygons, the additional terms in the series are very small due to the decay of the Fourier coefficients, and we suggest taking $N_1=20$ and $N_2=10$ for good accuracy.

\begin{table}
\centering
\def~{\hphantom{0}}
\begin{tabular}{ c  c  c }
 Shape & $N_1$ & $N_2$ \\ 
 \hline 
 Triangle &  2 & 1 \\ 
 Square &  2 & 1 \\
 Pentagon & 3 & 4 \\
 Hexagon & 3 & 5 
\end{tabular}
\caption{\label{tab:smoothing}Suitable upper limits for the truncation of the Fourier series representations for the evaporative flux for smoothed polygonal droplets \eqref{eq:JSol_Gen}. In all cases $N_3=N_2$.}
\end{table}

\begin{figure}
\begin{center}
\begin{tabular}{cc}
\includegraphics[width=0.25\textwidth]{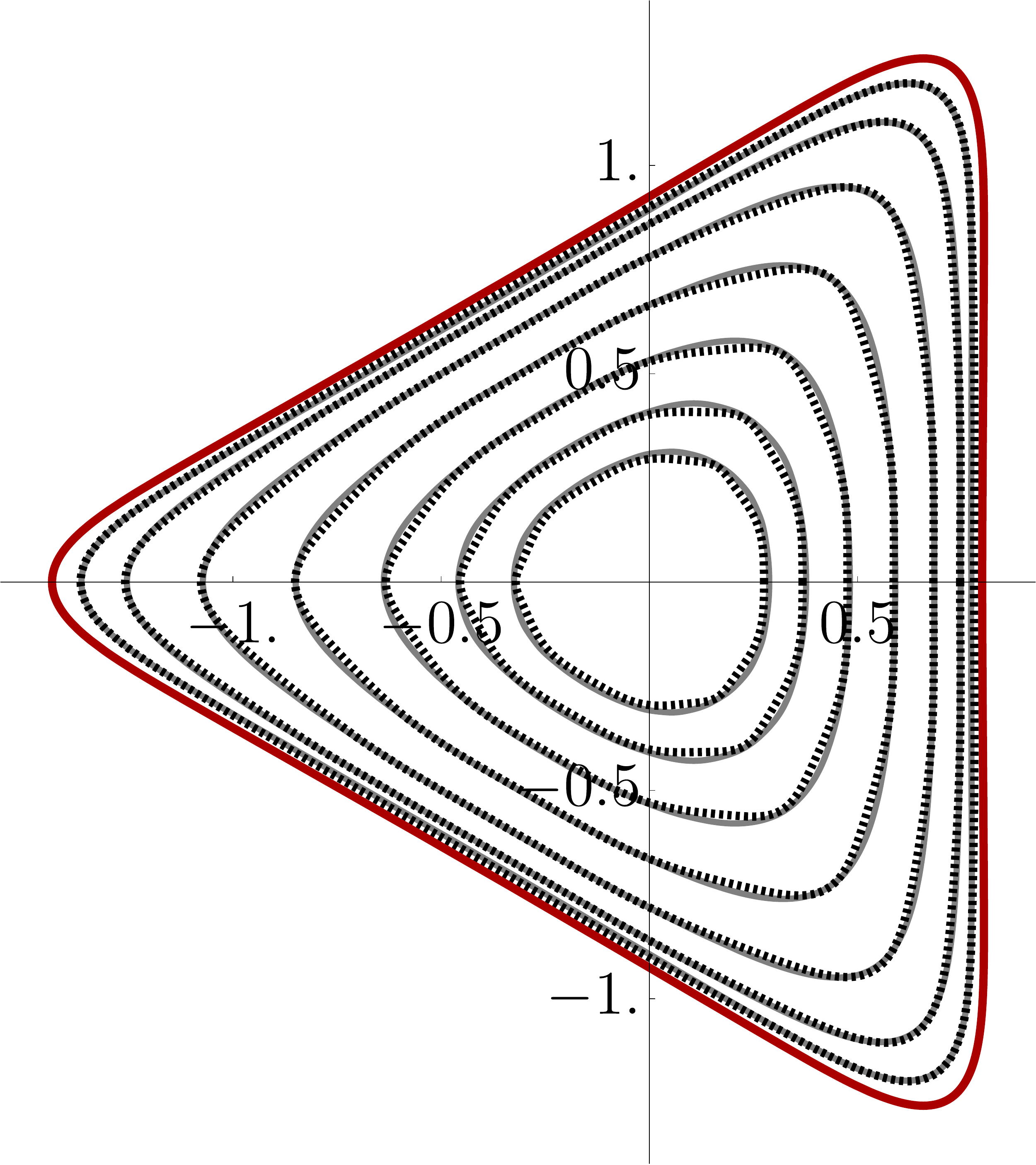}&
\includegraphics[width=0.45\textwidth]{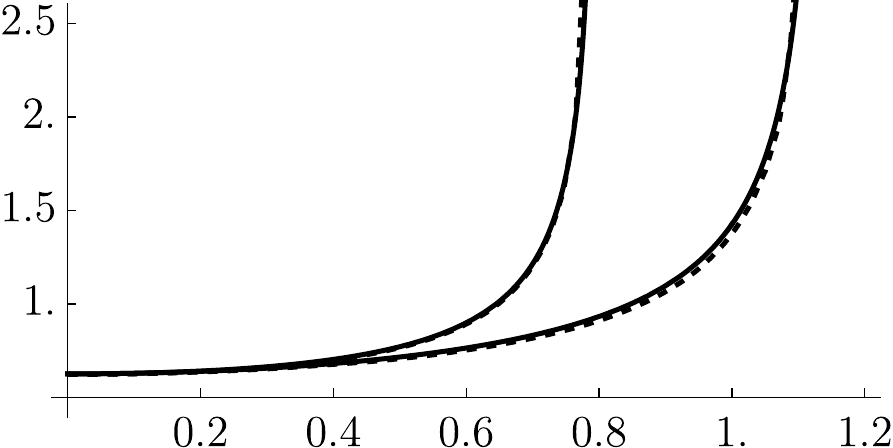}\\
(a) & (b) \\
\includegraphics[width=0.25\textwidth]{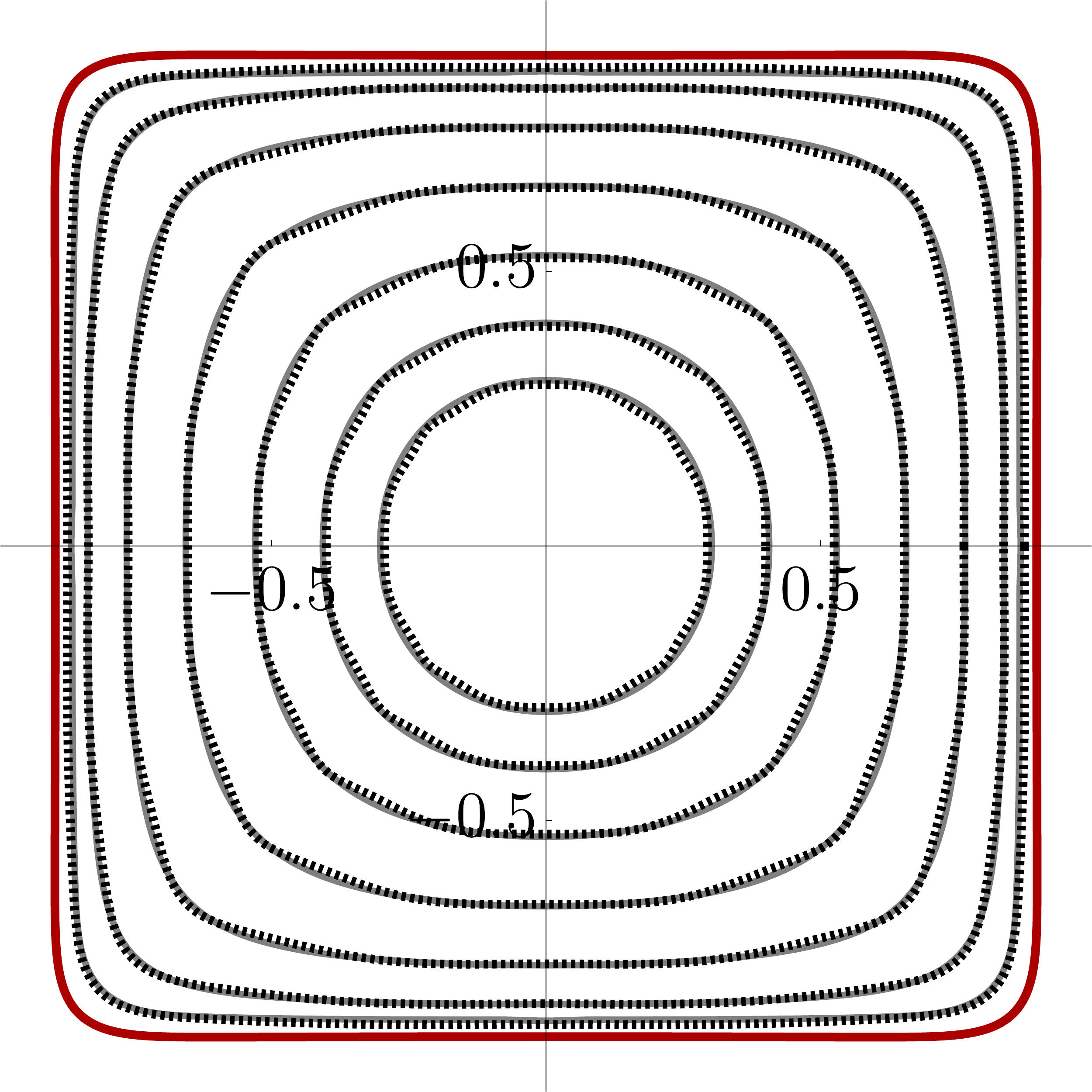}&
\includegraphics[width=0.45\textwidth]{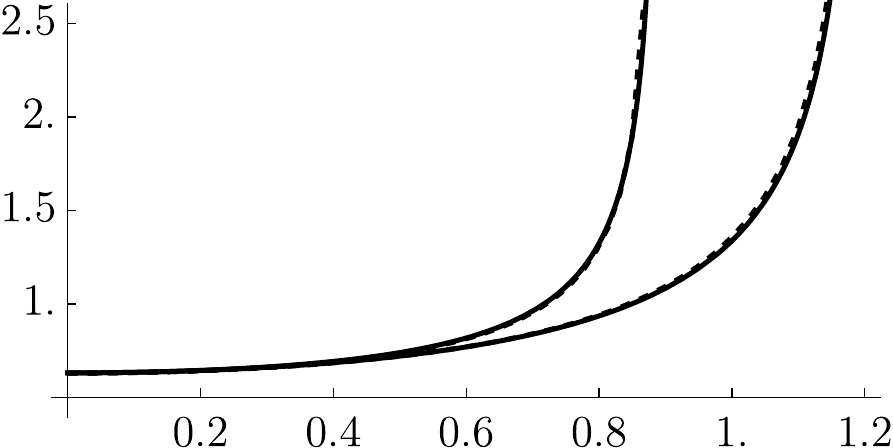}\\
(c) & (d) 
\end{tabular}
\end{center}
\caption{ (a), (c) Asymptotic (dashed) and numerical (solid, black) contours of evaporative fluxes; (b), (d) comparisons of asymptotic flux (solid curve) and numerical flux (dashed curve) for a smoothed triangular droplet (a), (b) and a smoothed square droplet (c), (d). In (a), (c), the red curves represent the pinned contact line.}
\label{fig:validationTriangleSquare}
\end{figure}

We plot the corresponding fluxes for triangular and square droplets in Figure \ref{fig:validationTriangleSquare} alongside numerical simulations of the full concentration problem (\ref{eq:Laplace})--(\ref{eq:Far_Field}). As with the monochromatic shapes, we see excellent agreement between the asymptotic and numerical results, even with the additional approximations discussed above. In particular, using the error metric defined in (\ref{eq:Error}), we find
\begin{align}
    e(3,0)&=9.6\times10^{-3},\\
    e(3,\pi/3)&=1.2\times10^{-2},\\
    e(4,0)&=1.1\times10^{-2},\\
    e(4,\pi/4)&=1.9\times10^{-2},
\end{align}
indicating strong agreement. These excellent comparisons give us encouragement to utilize the asymptotic results in our considerations of the internal flow dynamics and solute transport.

\subsubsection{Internal flow dynamics of large droplets with regular polygonal footprints}
\label{sec:Flow_Dyn_Polygons}

For droplets with polychromatic footprints, we set $N_2=0$ when modelling the internal flow dynamics so as to avoid the second-order complications. 
We shall see that this does not significantly diminish the resulting predictions. 
To this end, and following the model presented in \textsection \ref{sec:Flow_Dynamics}, we find
\begin{equation}
    \dd{h}{t}\approx-\pi/4, 
\end{equation}
and the liquid pressure has the expansion
\begin{equation}
 Q\approx Q_0(r)+\sum_{i=1}^N c_{ni}Q_1(r;ni)\cos ni\theta.
\end{equation}

\subsubsection{Residue from large droplets with regular polygonal footprints}
\label{sec:Residue_Polygons}

Once we have determined the pressure, the streamlines satisfy
\begin{equation}
    \psi=\theta_c+\psi_1,
\end{equation}
where
\begin{equation}
    \dd{\psi_1}{r}=\frac{p_\theta}{r^2p_r}\approx \frac{1}{Q_0(r)} \sum_{i=1}^N ni\,c_{ni}Q_1(r;ni)\sin ni\theta,
\end{equation}
which must be solved subject to
\begin{equation}
    \left.\psi_1\right|_{r=1}=0.
\end{equation}

In determining $Q$ and $\psi$, we must first expand the evaporative flux in a Fourier series. However, when expanding
\begin{equation}
    J=\frac{2}{\pi}\frac{a}{\sqrt{a^2-r^2}}\left\{1+\sum_{i=1}^{N}c_{ni} f_1(r;ni)\cos ni\theta \right\},
\end{equation}
there are two contributions involving $\cos ni \theta$: one from expanding the $a$ terms in the leading coefficient, and one from that multiplying $f_1$. While both of these contribute to $Q_1$, in principle the two should be summed separately: the former up to $N=\infty$ and the latter up to $N=N_1$ (in accordance with Table \ref{tab:smoothing}). However, it turns out that each separate contribution is substantially more complicated than using the two combined. This therefore suggests using two models: a ``simple" model where $N=N_1$, and an ``extended" model where $N=\infty$. We shall present solutions from both approaches in the sequel and discuss their accuracy and usefulness in reference to full numerical solutions. 

Once the pressure and streamlines have been found, we may finally determine the deposit density using
\begin{equation}
    D=\frac{1}{\sqrt{a^2+a_{\theta}^2}}\dd{M}{\theta_c}=\frac{1}{\sqrt{a^2+a_{\theta}^2}}\int_0^a r\q{\psi}{\theta_a}\ud r.
\end{equation}
In order to facilitate our comparisons, we note that all solutions for the density may be expanded as Fourier series of the form
\begin{equation}
    D=d_0+\sum d_i \cos ni\theta.\label{eq:compareResidueFourier}
\end{equation}
We present asymptotic and numerical calculations of the first few coefficients of this series in Table \ref{tab:polyCompare} for a number of different polygons. Both the simple and extended models are presented for the asymptotic results, while we give numerical results for both smoothed and sharp polygons. Throughout, we see that the asymptotic predictions do a remarkably good job of capturing the coefficients, particularly given that the expression for the evaporative flux was derived in the limit of nearly-circular droplets. Finally, we note that, in practice it was found that a mean of the simple model and the extended model gave the best agreement with DNS; we term this the ``averaged model", and use it for the remaining comparisons.

\begin{figure}
\begin{center}
\begin{tabular}{cc}
\includegraphics[width=0.3\textwidth]{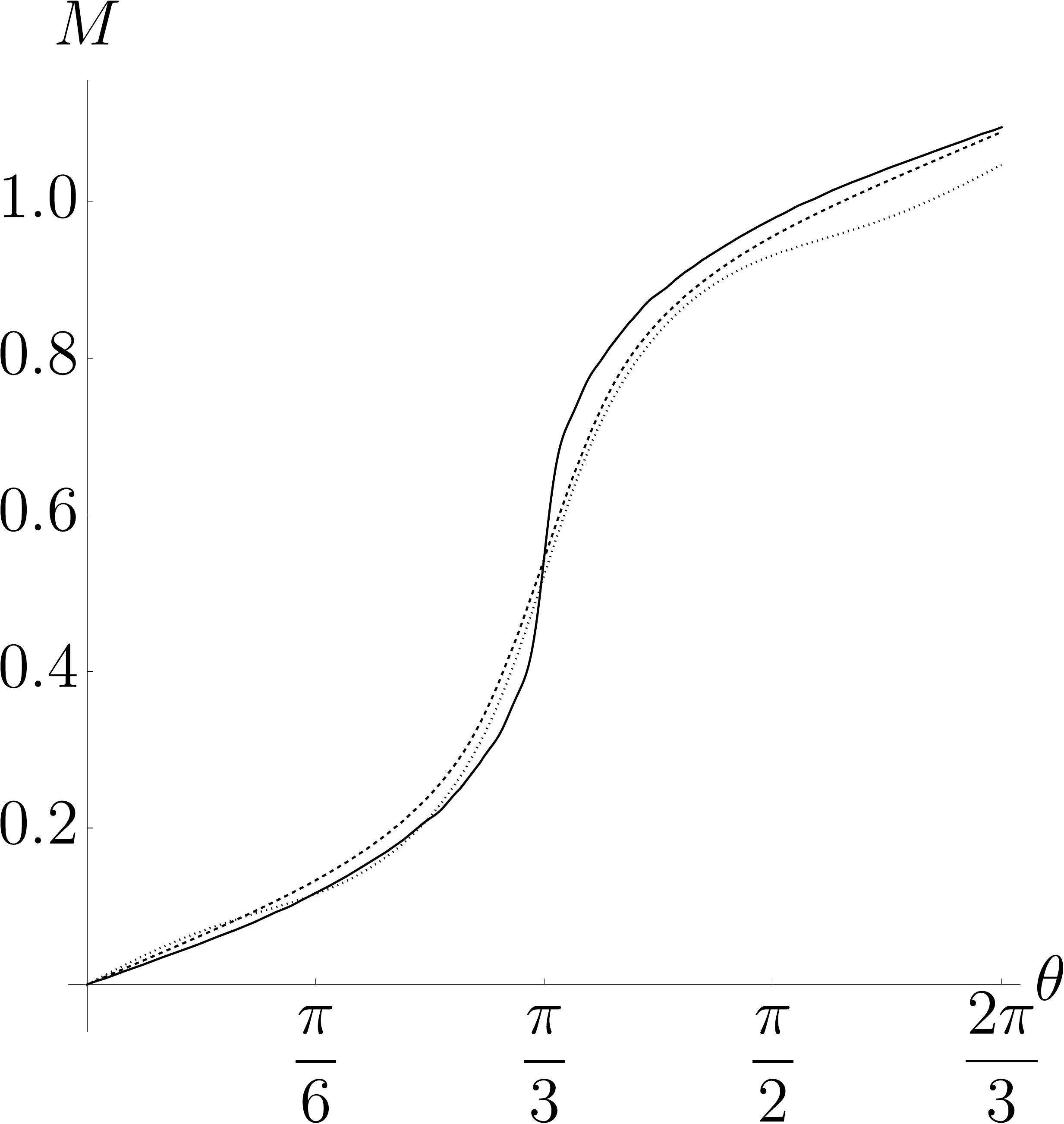}&
\includegraphics[width=0.3\textwidth]{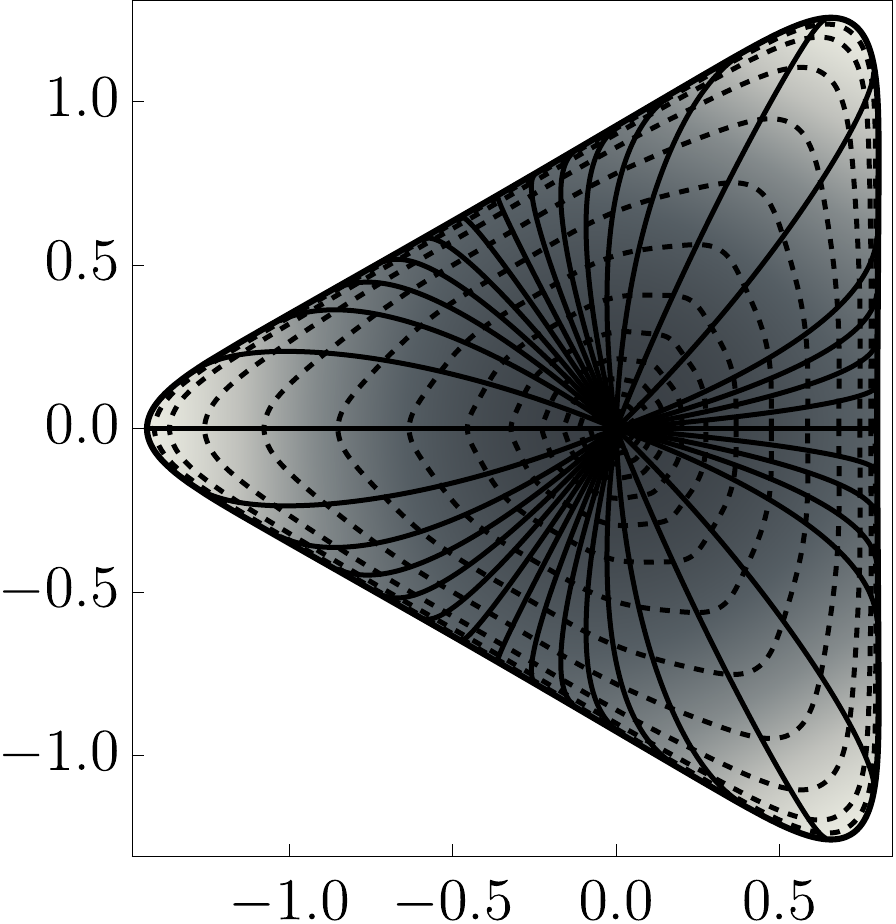}\\
(a) & (b) \\
\includegraphics[width=0.35\textwidth]{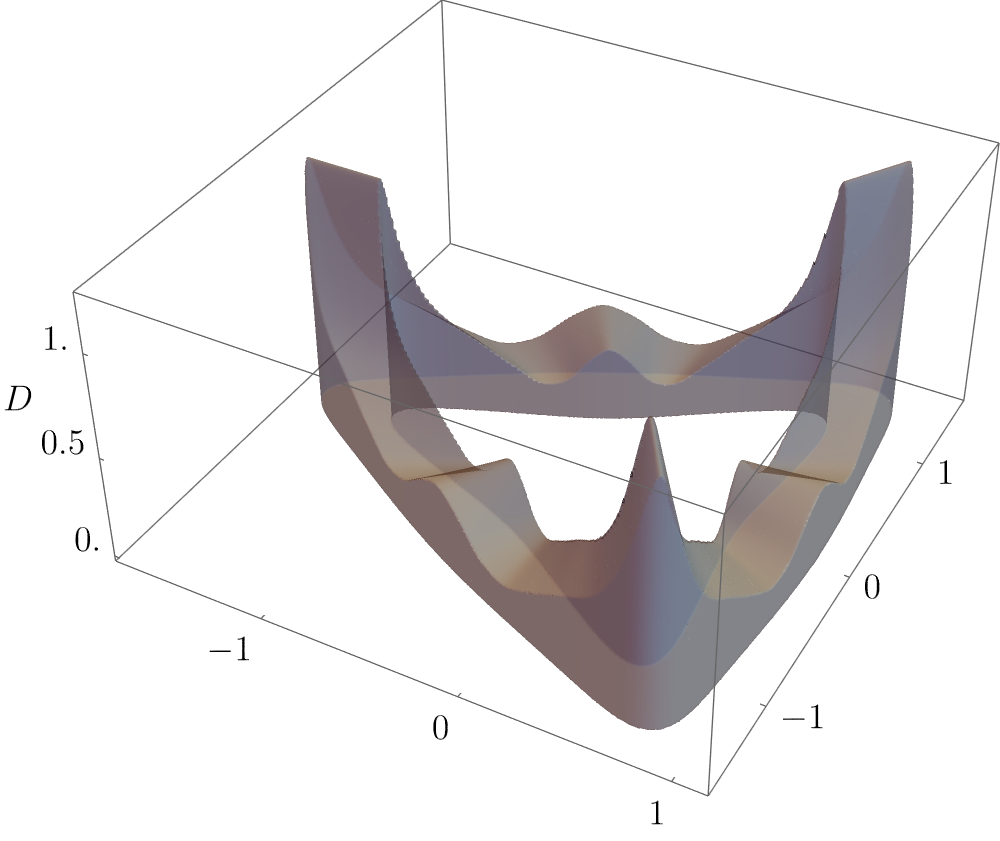}&
\includegraphics[width=0.35\textwidth]{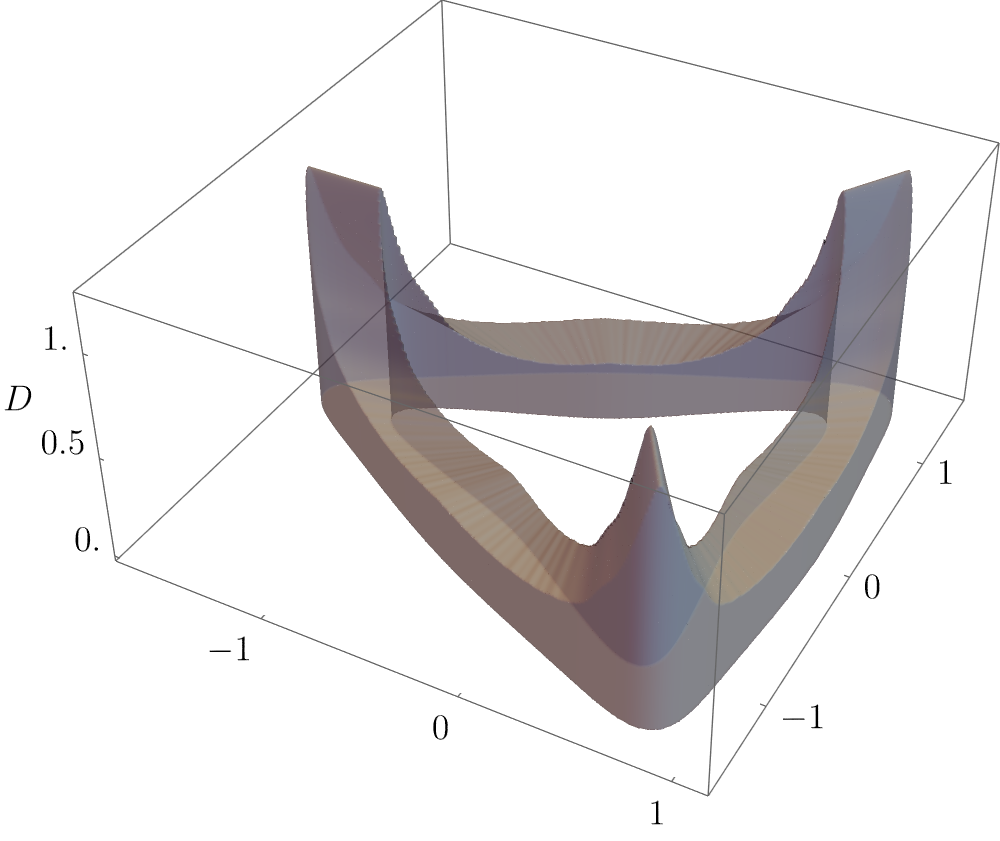}\\
(c) & (d) \\
\includegraphics[width=0.3\textwidth]{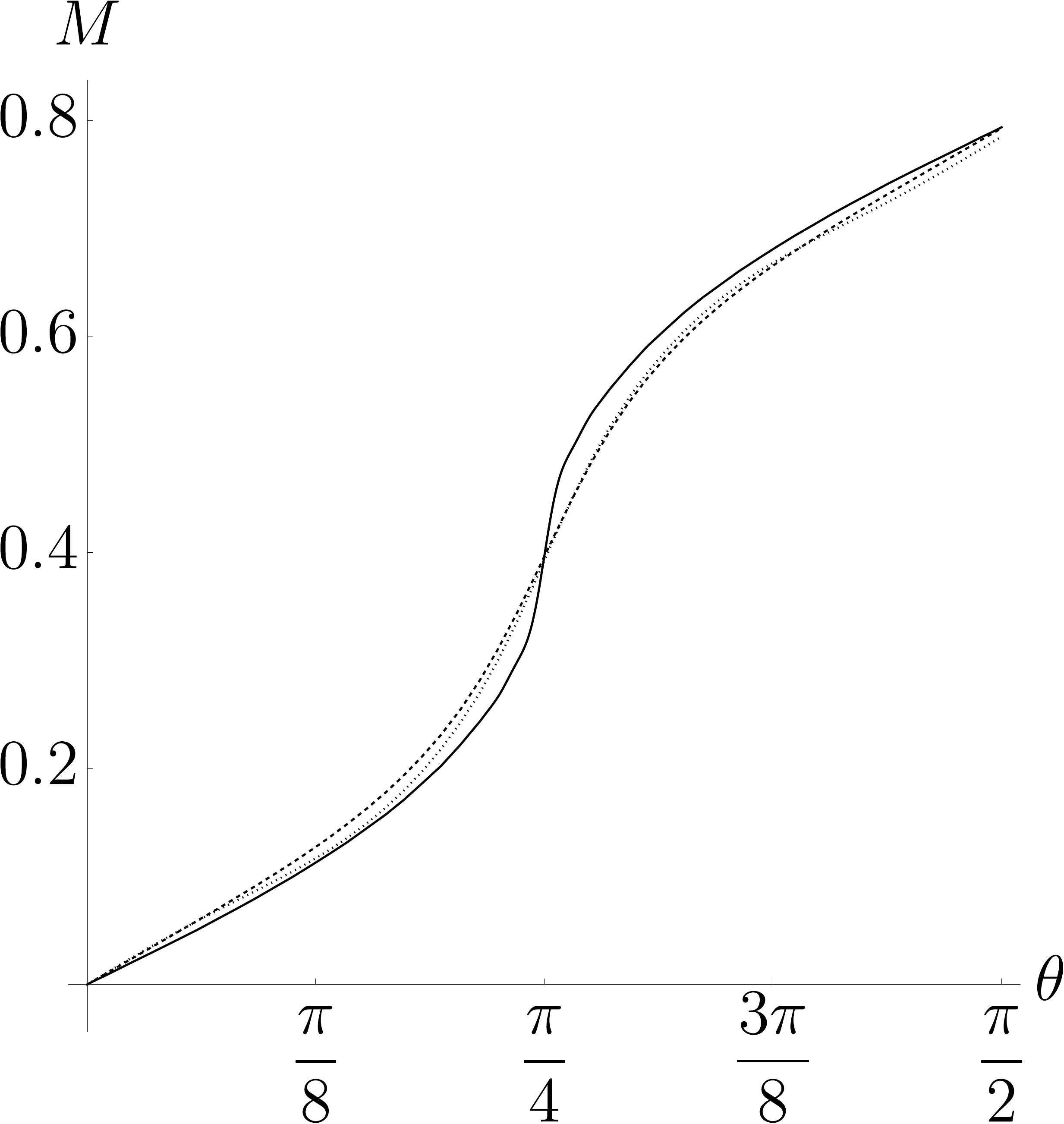}&
\includegraphics[width=0.3\textwidth]{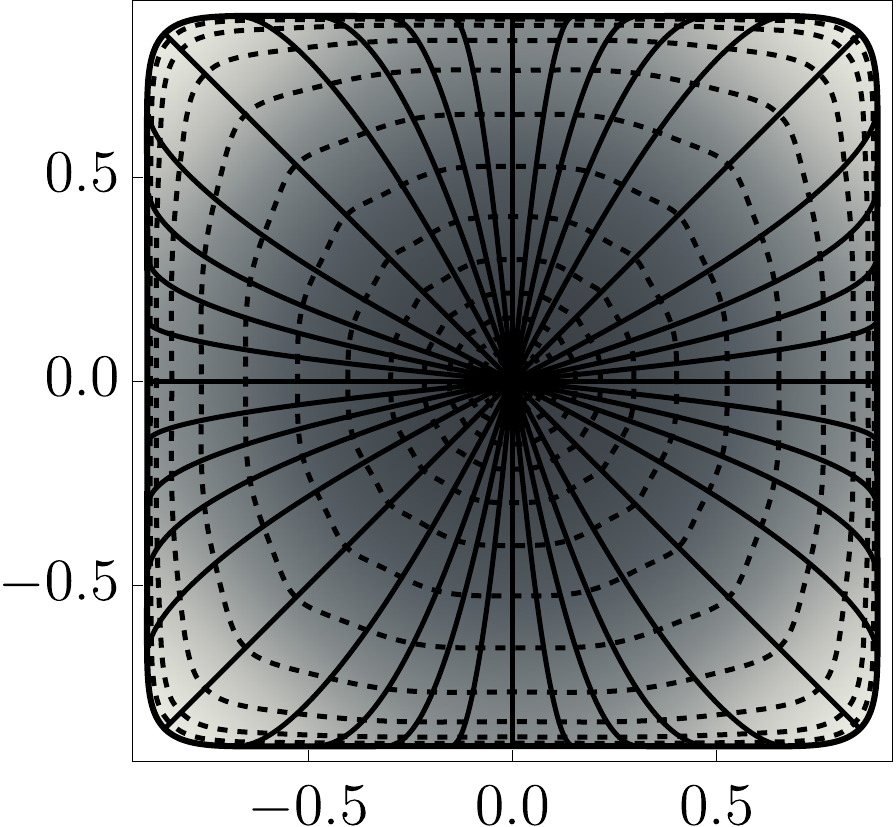}\\
(e) & (f) \\
\includegraphics[width=0.35\textwidth]{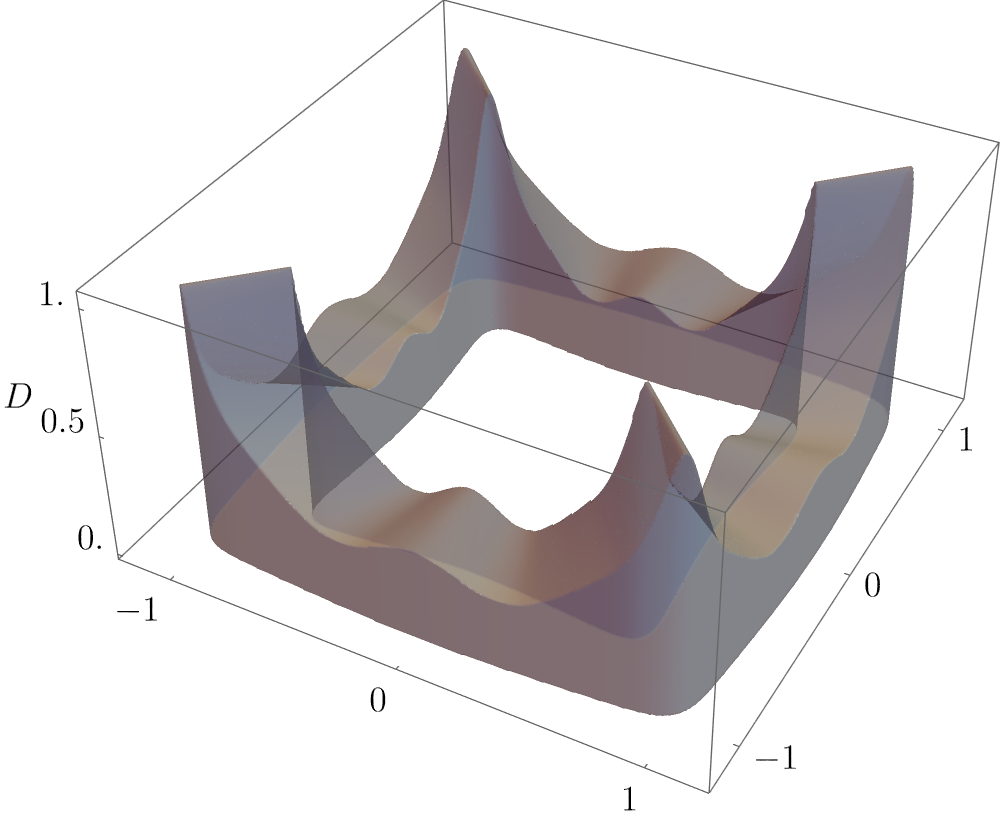}&
\includegraphics[width=0.35\textwidth]{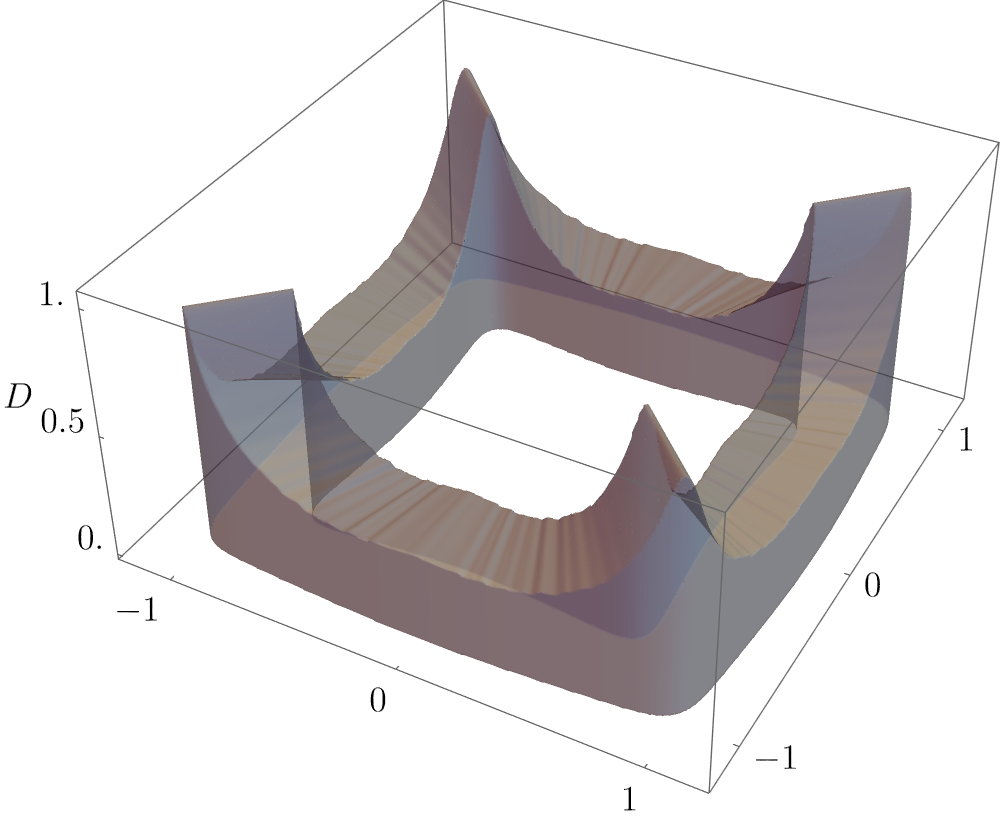}\\
(g) & (h)
\end{tabular}
\end{center}
\caption{Comparison plots for triangles (a)--(d) and squares (e)--(f), showing (a), (e) mass accumulation, with sharp-polygon DNS as a solid curve, smoothed-polygon DNS as a dashed curve, and averaged asymptotic model as a dotted curve; (b), (e) pressure (background colour), streamlines (solid curves) and contours of equal evaporative flux (dashed curves); (c), (g) residue density according to the averaged asymptotic model; (d), (h) the residue according to the smoothed DNS model.}
\label{fig:triSqResidue}
\end{figure}

\begin{figure}
\begin{center}
\begin{tabular}{cc}
\includegraphics[width=0.3\textwidth]{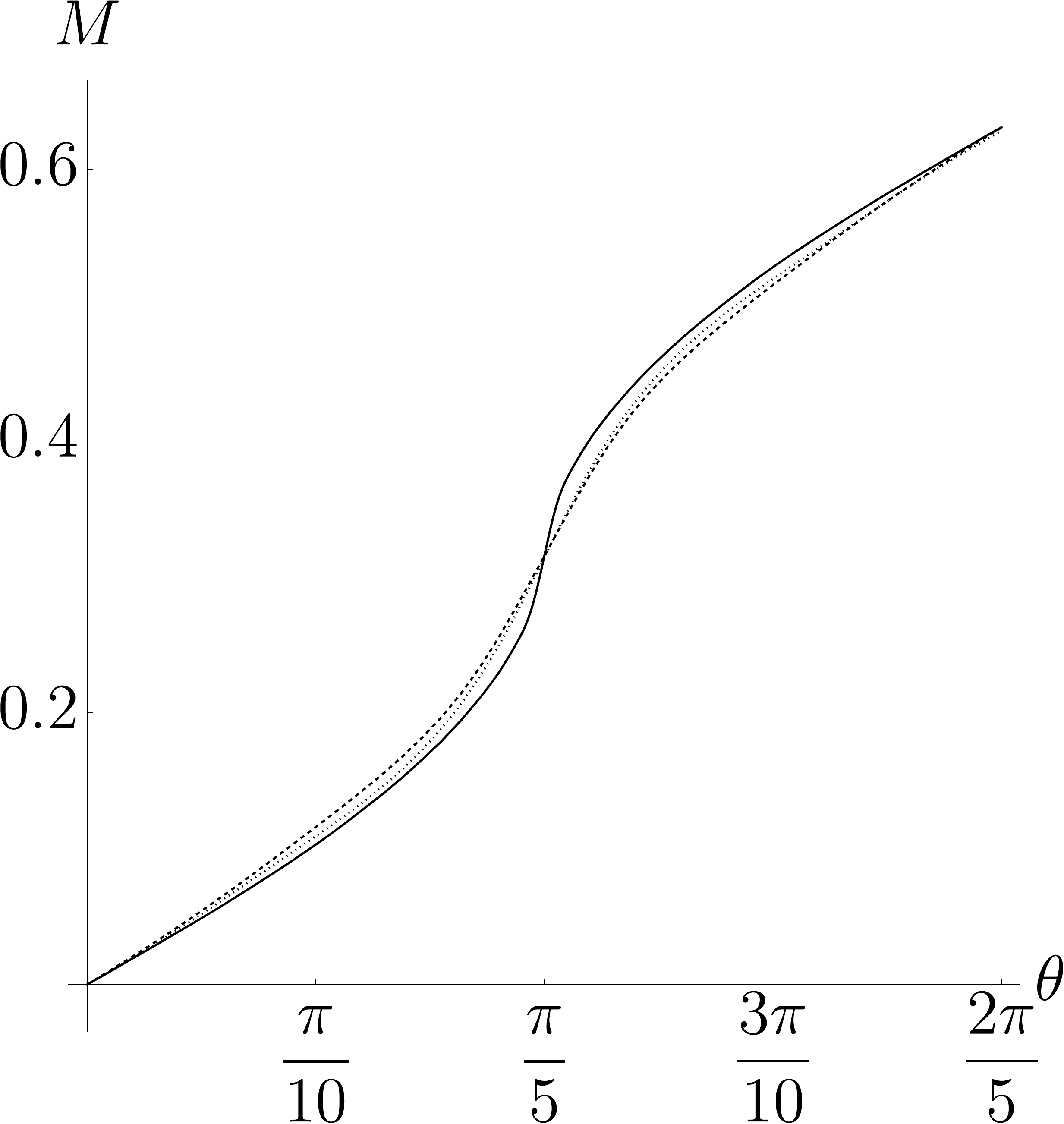}&
\includegraphics[width=0.3\textwidth]{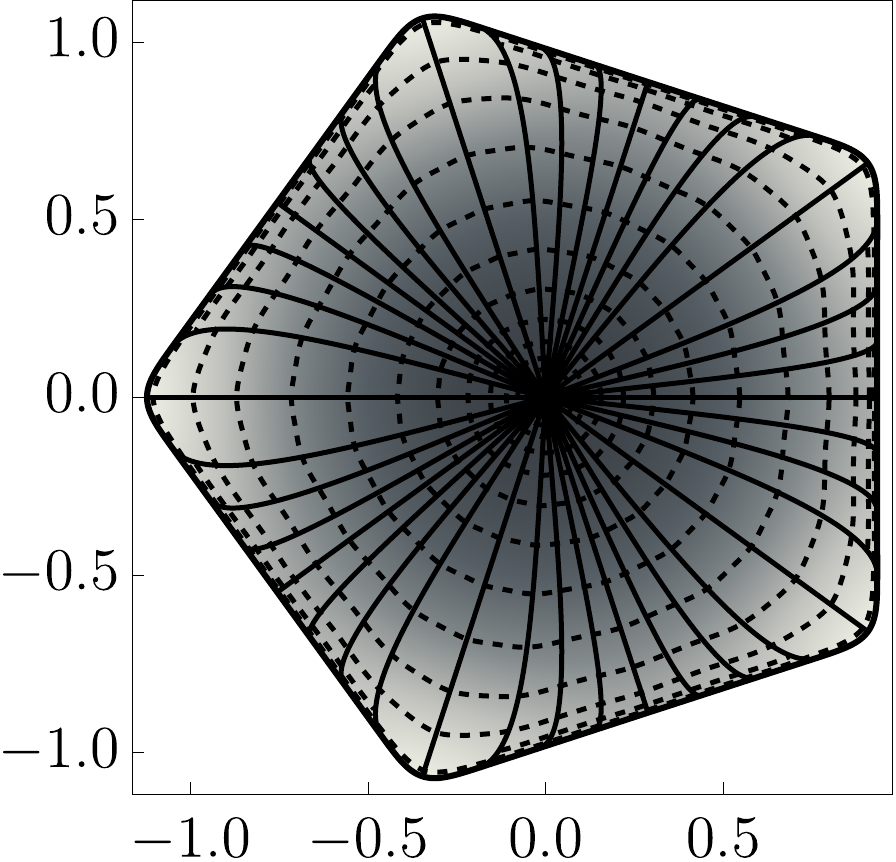}\\
(a) & (b) \\
\includegraphics[width=0.35\textwidth]{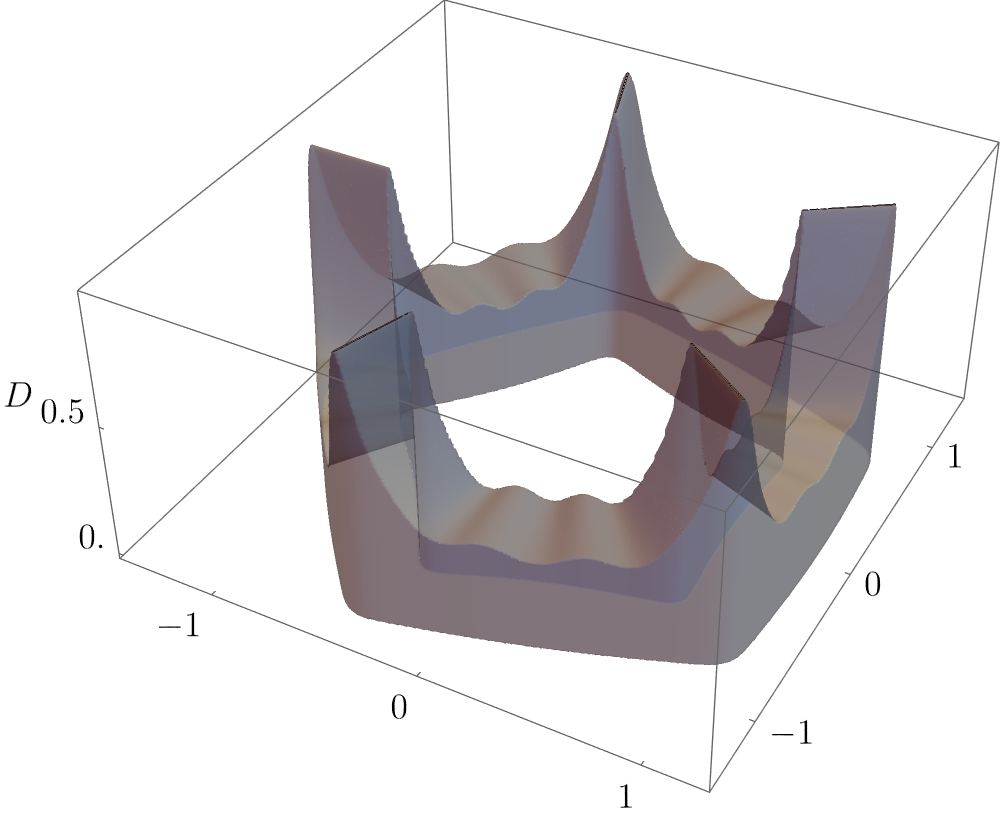}&
\includegraphics[width=0.35\textwidth]{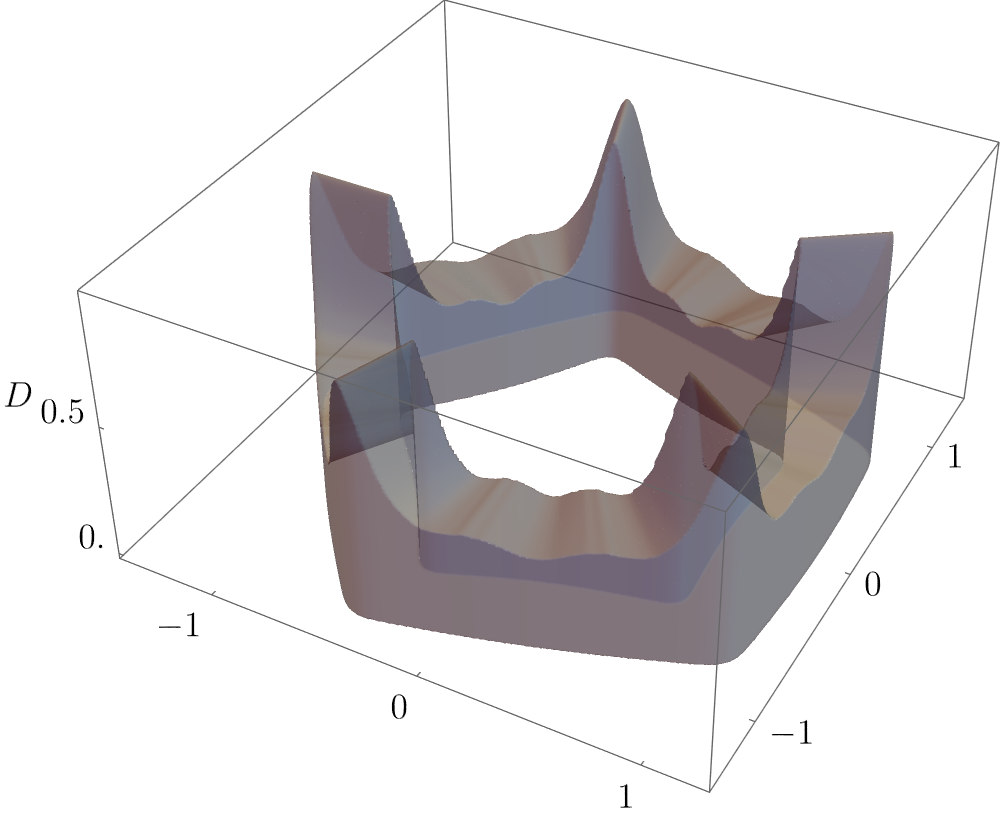}\\
(c) & (d) \\
\includegraphics[width=0.3\textwidth]{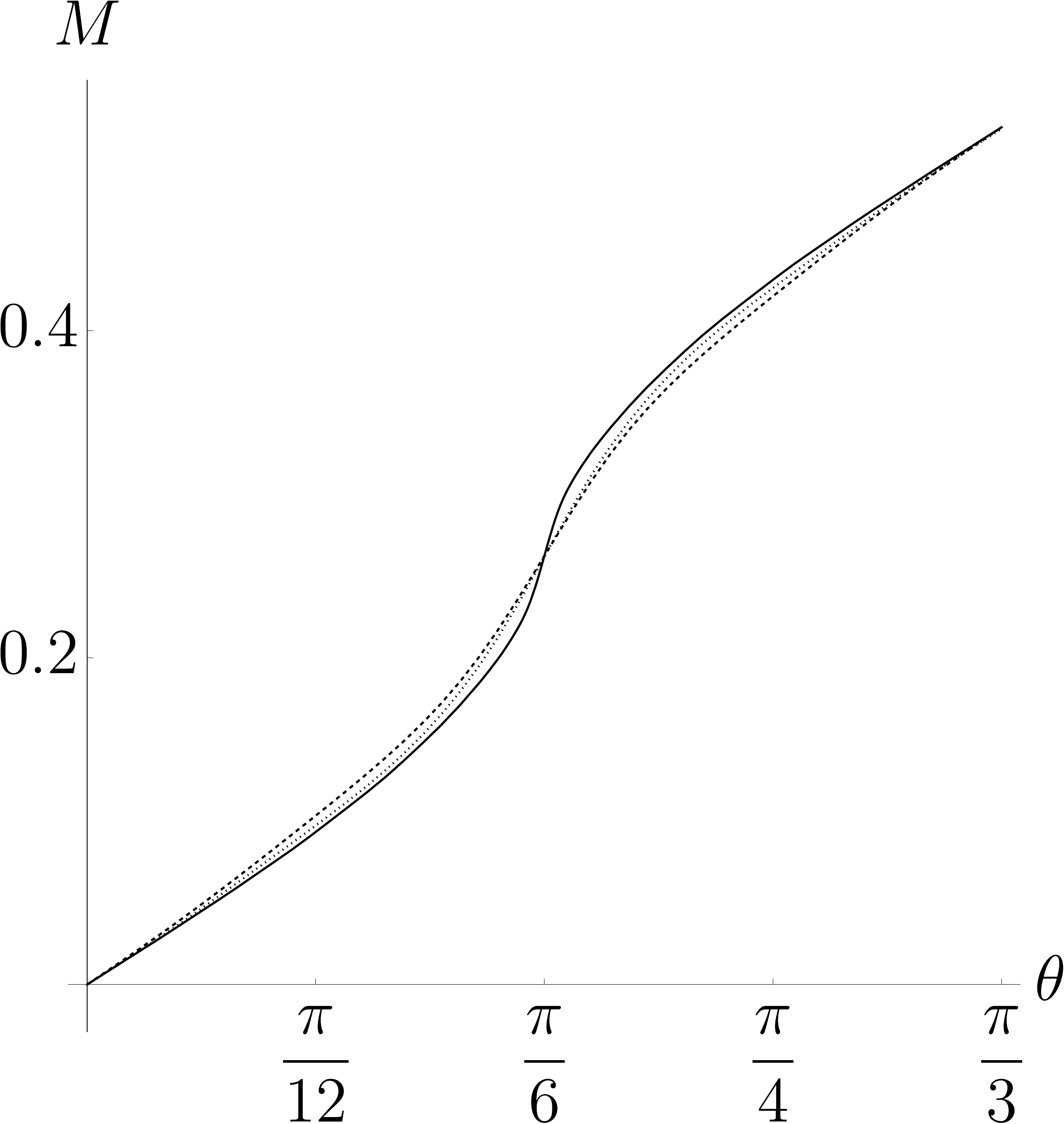}&
\includegraphics[width=0.3\textwidth]{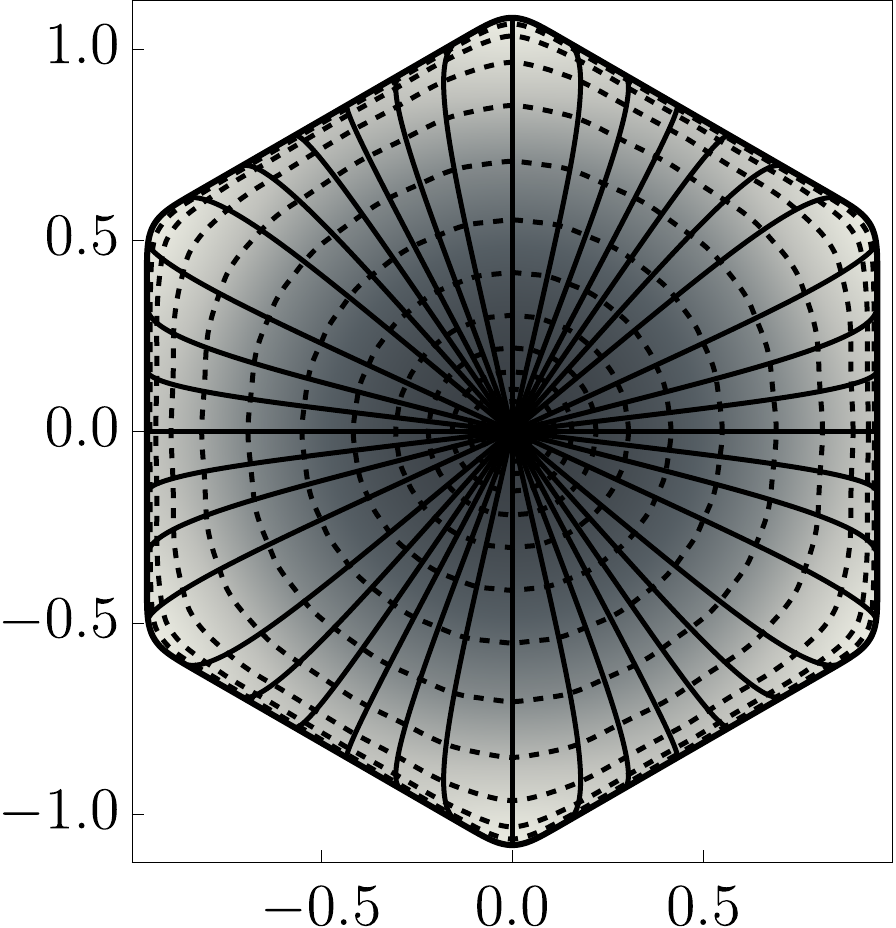}\\
(e) & (f) \\
\includegraphics[width=0.35\textwidth]{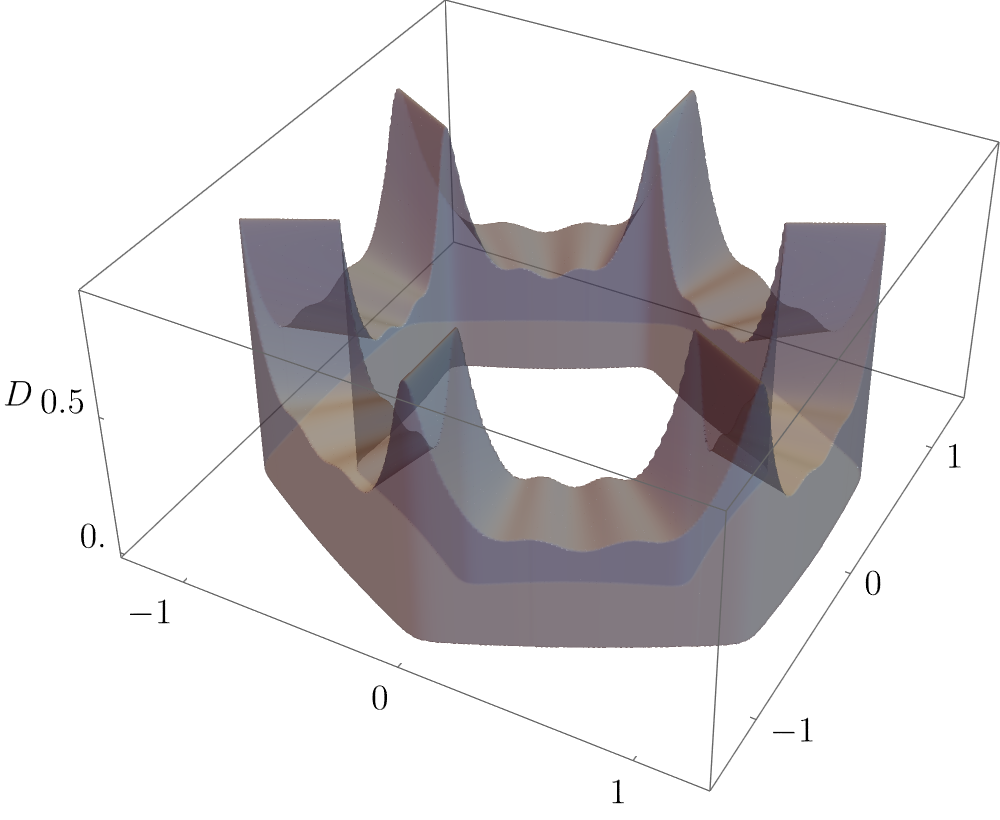}&
\includegraphics[width=0.35\textwidth]{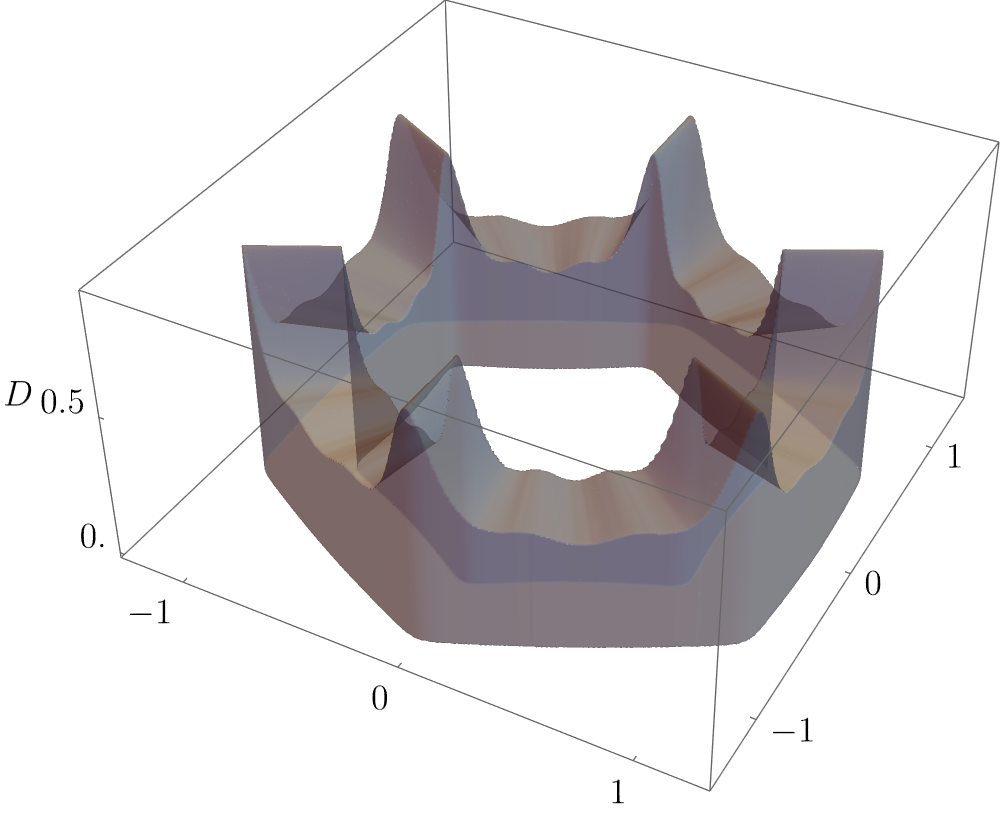}\\
(g) & (h)
\end{tabular}
\end{center}
\caption{As in Figure \ref{fig:triSqResidue}, for pentagons (a)--(d) and hexagons (e)--(f).}
\label{fig:penHexResid}
\end{figure}

We give comparisons between the various models for triangles and squares in Figure \ref{fig:triSqResidue} and for pentagons and hexagons in Figure \ref{fig:penHexResid}. In each case, the mass accumulated between $\theta=0$ and $\theta=2\pi/n$ (i.e.\, for the first period) is plotted for the respective $n$-gon for the DNS for the sharp $n$-gon (solid line), the DNS for the smoothed $n$-gon (dashed line) and the averaged model (dotted line). In all cases the smoothed-polygon DNS and the averaged model agree well, essentially obscuring one another for $n\geq 4$. For $n=3$ there is a small deviation close to the corner of the droplet. In all cases the sharp polygon demonstrates a significantly sharper increase in mass close to the corner than for the smoothed polygon.

The second plot for each shape shows the pressure (coloured background), streamlines (solid line) and contours of evaporative flux (dashed lines). The pressure is highest at the centre and lowest at the corners, driving a flow that is predominantly radially outwards, and preferentially towards the corners, hence the resulting streamlines. As anticipated, the contours of the evaporative flux are approximately circular close to the centre, and better approximate the shape of the contact line further out, as the effects of geometry become more pronounced.

The final plots show predicted residue densities according to the averaged asymptotic model and the smoothed DNS model. Again, the agreement is generally quite good, although the triangle and square models show some disagreement along the straight sides. Although the curvature is essentially zero here, the residue is still non-zero, indicating that the residue density is not solely due the curvature of the contact line.

Finally, we note an important detail about the streamlines. In these cases, the streamlines tend to converge towards the corners, as this is where the most liquid mass is being lost due to evaporation. This is in contrast to the case where the system is surface tension dominated. For example, Figure 3(e) of \citet{saenz2017dynamics} shows the streamlines diverging close to the corners. This is due to the effect of the interfacial height approaching zero there. We anticipate that in our problem, similar behaviour would be observed in the small, capillarity-induced boundary layer near the contact line that is not resolved here, where our primary goal was to illustrate the usefulness of the solution for the evaporative flux given by (\ref{eq:JSol}) and (\ref{eq:JSol_Gen}) in a specific application. Unfortunately, this means that further analysis of this boundary layer would be necessary to facilitate comparisons with experimental results such as those in \citet{saenz2017dynamics} to the residues predicted here. However, with $J(r,\theta)$ to hand, this is now something that can be readily addressed in future studies.

\begin{table}
\centering
\def~{\hphantom{0}}
\begin{tabular}{ c  c  c  c c c c }
 Method & $d_0$ & $d_1$ & $d_2$ & $d_3$ & $d_4$ & $d_5$ \\ 
  \hline 
 \multicolumn{6}{c}{Triangle}\\
 \hline 
 Smoothed numerical & $0.389845$ & $-0.185761$ & $0.152323$ & $-0.102641$ & $0.0765863$ & $-0.0551933$ \\
 Simple Approximate & $0.365957$ & $-0.188585$ & $0.234613$ & $0.0156282$ & $0.0322937$ & $-0.0327363$ \\
 Extended Approximate & $0.371964$ & $-0.19741$ & $0.204035$ & $-0.151659$ & $0.113871$ & $-0.0858848$ \\
 Sharp numerical & $0.333647$ & $-0.148966$ & $0.114942$ & $-0.100272$ & $0.0925472$ & $-0.0794294$\\
 \hline 
 \multicolumn{6}{c}{Square}\\
 \hline 
 Smoothed numerical & $0.442415$ & $-0.171044$ & $0.115701$ & $-0.0636093$ & $0.0439018$ & $-0.0298748$ \\
 Simple Approximate & $0.433424$ & $-0.196197$ & $0.161807$ & $-0.00288329$ & $0.0193128$ & $-0.018677$ \\
 Extended Approximate & $0.435694$ & $-0.199382$ & $0.152589$ & $-0.107735$ & $0.0770608$ & $-0.0550148$ \\
 Sharp numerical & $0.409369$ & $-0.190932$ & $0.140932$ & $-0.116876$ & $0.101247$ & $-0.0896022$ \\
  \hline 
 \multicolumn{6}{c}{Pentagon}\\
 \hline 
 Smoothed numerical & $0.464833$ & $-0.153297$ & $0.0954394$ & $-0.0778922$ & $0.0346092$ & $-0.0200515$ \\
 Simple Approximate & $0.460645$ & $-0.183146$ & $0.123048$ & $-0.0874442$ & $0.00728663$ & $-0.0116989$ \\
 Extended Approximate & $0.461384$ & $-0.184498$ & $0.123788$ & $-0.0844159$ & $0.0587529$ & $-0.0407542$ \\
 Sharp numerical & $0.44519$ & $-0.191098$ & $0.133475$ & $-0.109293$ & $0.0938368$ & $-0.082076$ \\
 \hline
  \multicolumn{6}{c}{Hexagon}\\
 \hline 
 Smoothed numerical & $0.476887$ & $-0.132762$ & $0.0791473$ & $-0.061533$ & $0.0261548$ & $-0.0144765$ \\
 Simple Approximate & $0.474262$ & $-0.167634$ & $0.103962$ & $-0.07047$ & $0.00588958$ & $-0.00794147$ \\
 Extended Approximate & $0.474668$ & $-0.168374$ & $0.104359$ & $-0.0688598$ & $0.0464603$ & $-0.031108$ \\
 Sharp numerical & $0.464299$ & $-0.178226$ & $0.121561$ & $-0.0949031$ & $0.0800684$ & $-0.0694642$ \\
 \hline
\end{tabular}
\caption{\label{tab:polyCompare}
Fourier coefficients according to the form \eqref{eq:compareResidueFourier} according to numerical calculations for the smoothed polygon, the simple approximate model, the extended approximate model, and numerical calculations for the sharp polygon.
}
\end{table}

\subsection{Droplets with other polygonal footprints}\label{sec:otherShapes}
\begin{figure}
\begin{center}
\begin{tabular}{cc}
\includegraphics[width=0.3\textwidth]{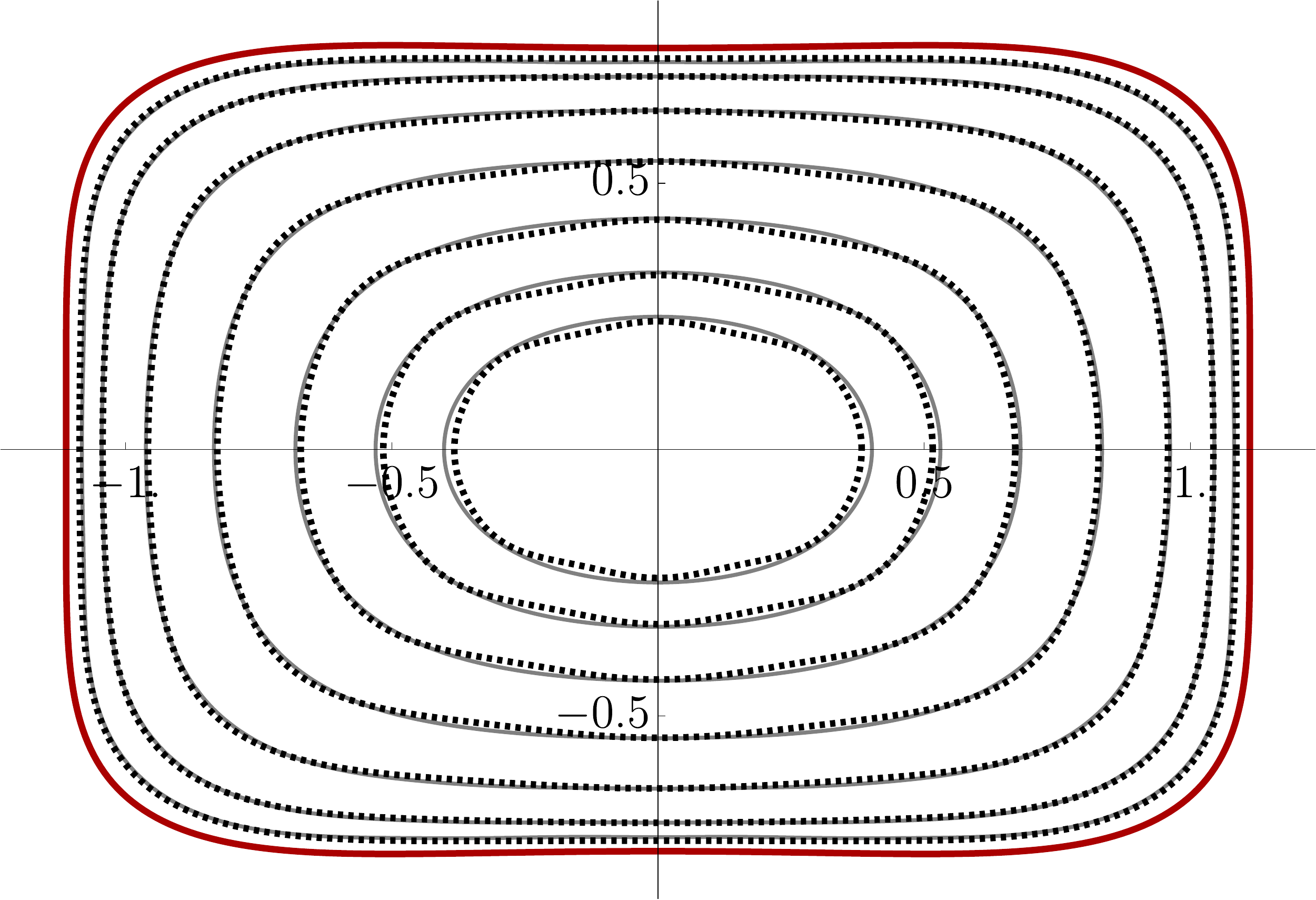}&
\includegraphics[width=0.3\textwidth]{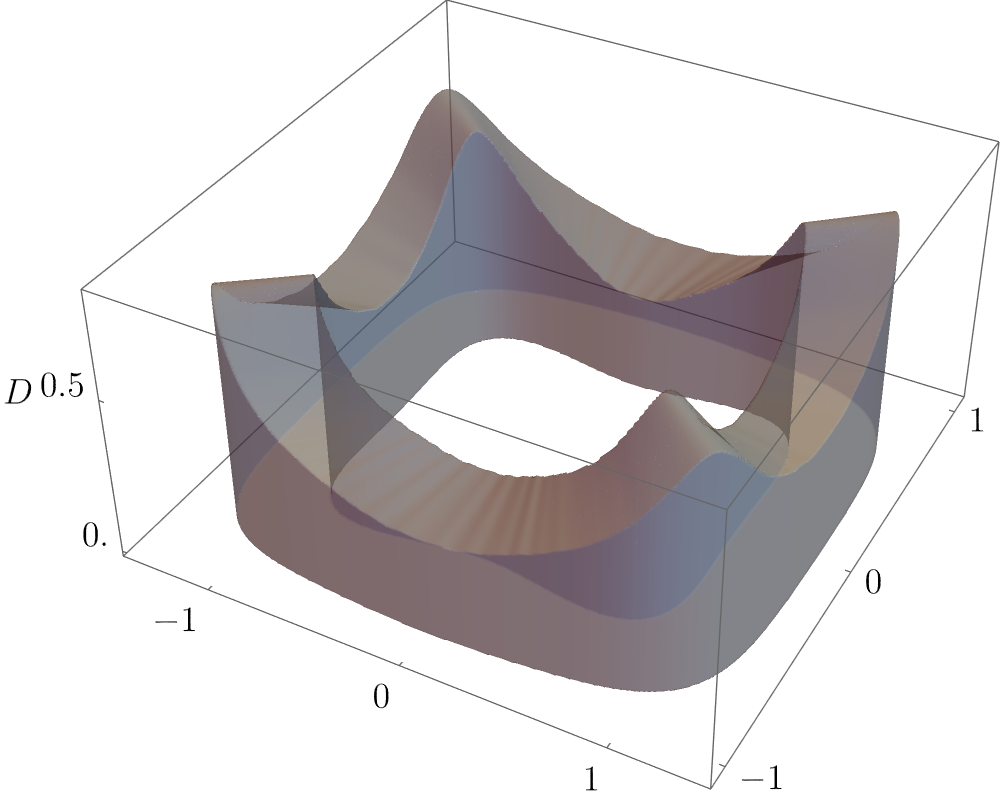}\\
(a) & (b) \\
\includegraphics[width=0.35\textwidth]{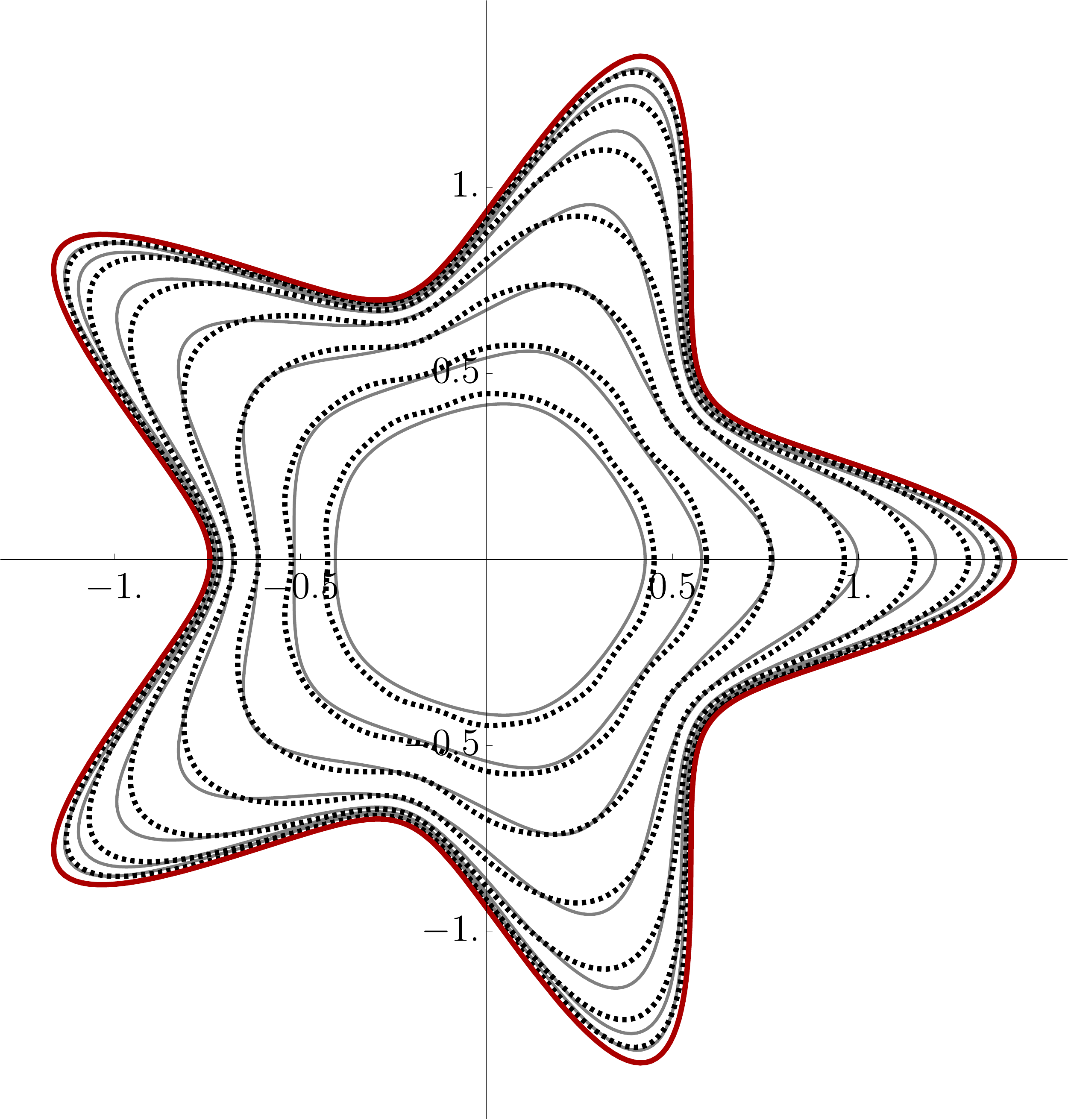}&
\includegraphics[width=0.35\textwidth]{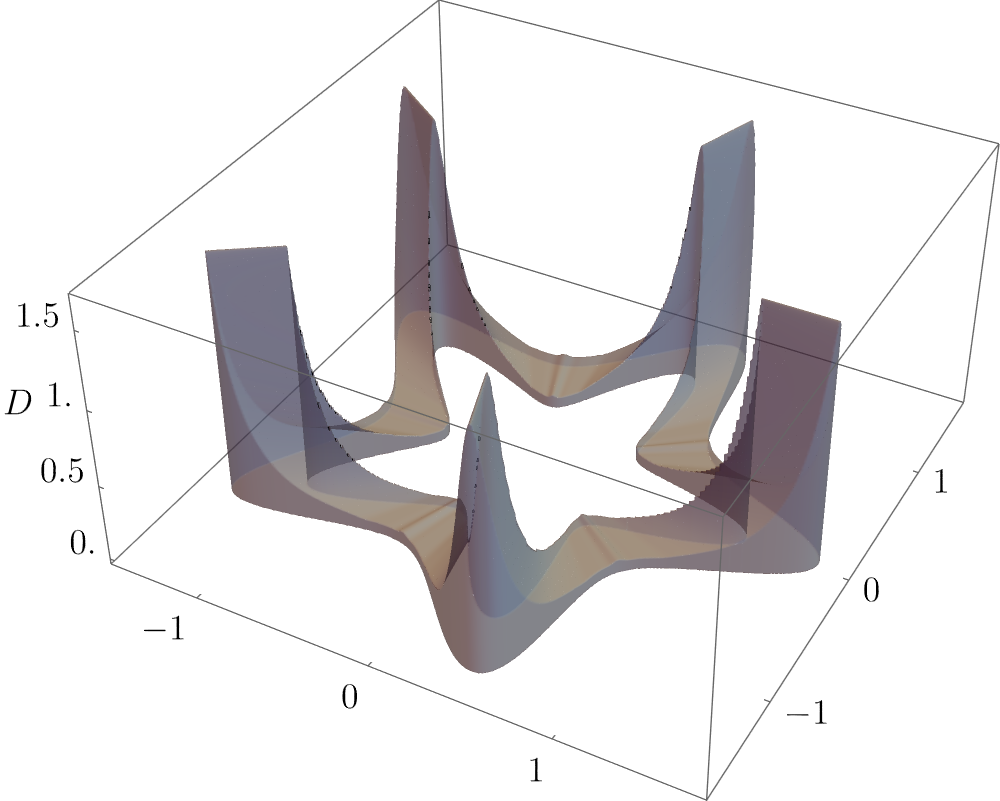}\\
(c) & (d) 
\end{tabular}
\end{center}
\caption{(a), (c)  Asymptotic (dashed curves) and numerical (solid,black curves) contours of evaporative flux and (b), (d) residues for (a), (b) a rectangle and (c), (d) a five-pointed star. The red curve denotes the pinned contact line of the droplet.} 
\label{fig:otherPolyResults}
\end{figure}

We finish by examining two further polygons. The first is a rectangle, selected as one of the simplest polygons which is not regular. The second is a five-pointed star, selected as a strongly non-convex polygon. Their evaporative fluxes and residues are given in Figure \ref{fig:otherPolyResults}.

One interesting note is that, despite having the same curvature, the residue for the rectangle at $\theta=0$ is higher than that at $\theta=\pi/2$ by $30.6\%$ (for comparison, the contact line at $\theta=0$ is $47.4\%$ further away than at $\theta=\pi/2$). It is noted that for higher aspect ratios, the agreement between asymptotic solutions and DNS became weaker, especially close to the origin. It is anticipated that this could be alleviated at least in part via the use of Galin's theorem \citep{sneddon1966mixed}.

We note that for the star, despite high levels of curvature and strong non-convexity, good agreement is still found for the contours of the evaporative flux, indicating that, at least for smoothed polygons, this method is quite robust. What is more, we actually have large areas of {\em negative} curvature here. As expected, the residue is very low in these areas.

Finally, we note that the contours close to the centre of the droplet are approximately circles for the five-pointed star, and ellipses for the rectangle. This is in stark contrast to the solution predicted by the leading order solution of \citet{fabrikant1986capacity}, which would predict star-like and rectangular contours all the way to the origin. This highlights the necessity and accuracy of the corrections to the evaporative flux derived herein.

\section{Conclusions}
\label{sec:Conclusions}
In the present work, we have examined the evaporation of non-circular droplets in the diffusion-limited regime. While the initial asymptotic analyses were based on the assumption that the droplet was nearly circular, comparisons against numerical simulations have shown good agreement far outside this regime. This has been tested for particular cases of both industrial and scientific interest, including polygons, strongly non-convex shapes, and irregular polygons. Given the paucity of analytic solutions for the evaporation of droplets, we anticipate that this model will be of significant interest to the droplet community, opening up numerous previously inaccessible avenues. We have demonstrated one of these by elucidating the final pattern of solute deposition from large, thin, particle-laden droplets with results supporting previous studies that demonstrate increased deposition at high-curvature parts of the contact line \cite[][]{FreedBrown2015,saenz2017dynamics,moore2022nascent}. 

Finally, we anticipate that, as the results in this paper are fundamentally new results in potential theory, they will be of interest to researchers in fields such as elasticity, thermodynamics, contact mechanics and electrostatics, among many others.

\appendix

\section{Numerical methods}\label{sec:appNumMeth}

\subsection{Evaporative flux}

Numerical evaporative fluxes were determined by solving (\ref{eq:Laplace})--(\ref{eq:Far_Field}) using COMSOL multiphysics. Under the thin assumption, the droplet was implemented as a two-dimensional surface of the appropriate shape, on which the Dirichlet condition (\ref{eqn:Surface_Conds}) was imposed, while extending the problem symmetrically to $z<0$ guarantees that the Neumann condition on the non-wetted part of the substrate is satisfied automatically. The far-field behaviour \eqref{eq:Far_Field} was imposed by implementing a Dirichlet condition on a large sphere centred at the origin of dimensionless radius $R_{\mathrm{sph}}\gg1$. Extensive testing showed that the results were insensitive to the choice of sufficiently large $R_{\mathrm{sph}}$; we take $R_{\mathrm{sph}} = 500$ in all our simulations. 

\subsection{Liquid pressure}

Solving the thin-film equation (\ref{eq:ThinFilm}) subject to (\ref{eq:NoFlux}) for the liquid pressure is an under-constrained problem due to the Neumann boundary condition. While this can be resolved in theory via an additional integral constraint, in practice it was easier to use the method of false transients \citep{mallinson1973method}. This was implemented in both COMSOL and Mathematica for verification, with solutions for all droplets presented in this paper being in excellent agreement.

For the final steps of using the numerical pressure solution to compute masses and corresponding densities, we integrated between suitable streamlines directly in Mathematica.

\bibliographystyle{abbrvnat}
\bibliography{NonCircular}

\begin{thebibliography}{52}
\providecommand{\natexlab}[1]{#1}
\providecommand{\url}[1]{\texttt{#1}}
\expandafter\ifx\csname urlstyle\endcsname\relax
  \providecommand{\doi}[1]{doi: #1}\else
  \providecommand{\doi}{doi: \begingroup \urlstyle{rm}\Url}\fi

\bibitem[Argatov(2011)]{argatov2011electrical}
I.~Argatov.
\newblock Electrical contact resistance, thermal contact conductance and
  elastic incremental stiffness for a cluster of microcontacts: Asymptotic
  modelling.
\newblock \emph{Quar. J. Mech. Appl. Math.}, 64\penalty0 (1):\penalty0 1--24,
  2011.

\bibitem[Basi et~al.(2013)Basi, Hunsche, and Noga]{basi2013effects}
S.~Basi, M.~Hunsche, and G.~Noga.
\newblock Effects of surfactants and the kinetic energy of monodroplets on the
  deposit structure of glyphosate at the micro-scale and their relevance to
  herbicide bio-efficacy on selected weed species.
\newblock \emph{Weed Res.}, 53\penalty0 (1):\penalty0 1--11, 2013.

\bibitem[Boersma and Danick(1993)]{boersma1993solution}
J.~Boersma and E.~Danick.
\newblock On the solution of an integral equation arising in potential problems
  for circular and elliptic disks.
\newblock \emph{SIAM J. Appl. Math.}, 53\penalty0 (4):\penalty0 931--941, 1993.

\bibitem[Borodachev and Galin(1974)]{borodachev1974contact}
N.~Borodachev and L.~Galin.
\newblock Contact problem for a stamp with narrow rectangular base.
\newblock \emph{J. Appl. Math. Mech.}, 38\penalty0 (1):\penalty0 108--113,
  1974.

\bibitem[Brutin and Starov(2018)]{brutin2018recent}
D.~Brutin and V.~Starov.
\newblock Recent advances in droplet wetting and evaporation.
\newblock \emph{Chem. Soc. Rev.}, 47\penalty0 (2):\penalty0 558--585, 2018.

\bibitem[Choi et~al.(2010)Choi, Stassi, Pisano, and Zohdi]{Choi2010}
S.~Choi, S.~Stassi, A.~P. Pisano, and T.~I. Zohdi.
\newblock Coffee-ring effect-based three dimensional patterning of
  micro/nanoparticle assembly with a single droplet.
\newblock \emph{Langmuir}, 26\penalty0 (14):\penalty0 11690--11698, 2010.

\bibitem[Copson(1947)]{copson1947problem}
E.~Copson.
\newblock On the problem of the electrified disc.
\newblock \emph{Proc. Edin. Math. Soc.}, 8\penalty0 (1):\penalty0 14--19, 1947.

\bibitem[Deegan et~al.(1997)Deegan, Bakajin, Dupont, Huber, Nagel, and
  Witten]{deegan1997capillary}
R.~D. Deegan, O.~Bakajin, T.~F. Dupont, G.~Huber, S.~R. Nagel, and T.~A.
  Witten.
\newblock Capillary flow as the cause of ring stains from dried liquid drops.
\newblock \emph{Nature}, 389\penalty0 (6653):\penalty0 827--829, 1997.

\bibitem[Deegan et~al.(2000)Deegan, Bakajin, Dupont, Huber, Nagel, and
  Witten]{deegan2000contact}
R.~D. Deegan, O.~Bakajin, T.~F. Dupont, G.~Huber, S.~R. Nagel, and T.~A.
  Witten.
\newblock Contact line deposits in an evaporating drop.
\newblock \emph{Phys. Rev. E}, 62\penalty0 (1):\penalty0 756, 2000.

\bibitem[Du and Deegan(2015)]{du2015ring}
X.~Du and R.~Deegan.
\newblock Ring formation on an inclined surface.
\newblock \emph{J. Fluid Mech.}, 775, 2015.

\bibitem[Dunn et~al.(2008)Dunn, Wilson, Duffy, David, and
  Sefiane]{dunn2008mathematical}
G.~Dunn, S.~Wilson, B.~Duffy, S.~David, and K.~Sefiane.
\newblock A mathematical model for the evaporation of a thin sessile liquid
  droplet: Comparison between experiment and theory.
\newblock \emph{Coll. \& Surf. A}, 323\penalty0 (1-3):\penalty0 50--55, 2008.

\bibitem[Eales et~al.(2015)Eales, Routh, Dartnell, and
  Simon]{eales2015evaporation}
A.~D. Eales, A.~F. Routh, N.~Dartnell, and G.~Simon.
\newblock Evaporation of pinned droplets containing polymer--an examination of
  the important groups controlling final shape.
\newblock \emph{AIChE Journal}, 61\penalty0 (5):\penalty0 1759--1767, 2015.

\bibitem[Fabrikant(1985)]{fabrikant1985potential}
V.~Fabrikant.
\newblock On the potential flow through membranes.
\newblock \emph{ZAMP}, 36\penalty0 (4):\penalty0 616--623, 1985.

\bibitem[Fabrikant(1986)]{fabrikant1986capacity}
V.~Fabrikant.
\newblock On the capacity of flat laminae.
\newblock \emph{Electromagnetics}, 6\penalty0 (2):\penalty0 117--128, 1986.

\bibitem[Fabrikant(1987)]{fabrikant1987diffusion}
V.~Fabrikant.
\newblock Diffusion through perforated membranes.
\newblock \emph{J. Appl. Phys.}, 61\penalty0 (3):\penalty0 813--816, 1987.

\bibitem[Fabrikant(1991)]{fabrikant1991mixed}
V.~Fabrikant.
\newblock \emph{Mixed boundary value problems of potential theory and their
  applications in engineering}.
\newblock Springer, 1991.

\bibitem[Freed-Brown(2015)]{FreedBrown2015}
J.~E. Freed-Brown.
\newblock \emph{Deposition from evaporating drops: power laws and new
  morphologies in coffee stains}.
\newblock PhD thesis, University of Chicago, 2015.

\bibitem[Galin et~al.(1961)Galin, Moss, and Sneddon]{galin1961contact}
L.~A. Galin, H.~Moss, and I.~N. Sneddon.
\newblock Contact problems in the theory of elasticity.
\newblock Technical report, NC State Univ., Raleigh Sch. of Phys. Sci. and
  Appl. Math., 1961.

\bibitem[Guazzelli and Pouliquen(2018)]{Guazzelli2018}
{\'E}.~Guazzelli and O.~Pouliquen.
\newblock Rheology of dense granular suspensions.
\newblock \emph{J. Fluid Mech.}, 852:\penalty0 P1, 2018.

\bibitem[Harris et~al.(2007)Harris, Hu, Conrad, and Lewis]{Harris2007}
D.~J. Harris, H.~Hu, J.~C. Conrad, and J.~A. Lewis.
\newblock Patterning colloidal films via evaporative lithography.
\newblock \emph{Phys. Rev. Lett.}, 98\penalty0 (14):\penalty0 148301, 2007.

\bibitem[Hu and Larson(2002)]{Hu2002}
H.~Hu and R.~G. Larson.
\newblock Evaporation of a sessile droplet on a substrate.
\newblock \emph{J. Phys. Chem. B}, 106\penalty0 (6):\penalty0 1334--1344, 2002.

\bibitem[Huo et~al.(2020)Huo, Shao, Dong, Liang, Bi, He, Li, Gao, and
  Song]{huo2020real}
S.-T. Huo, L.-Q. Shao, T.~Dong, J.-S. Liang, Z.-T. Bi, M.~He, Z.~Li, Z.~Gao,
  and J.-Y. Song.
\newblock Real rgb printing amoled with high pixel per inch value.
\newblock \emph{J. Soc. for Inf. Disp.}, 28\penalty0 (1):\penalty0 36--43,
  2020.

\bibitem[Jackson(1999)]{jackson1999classical}
J.~D. Jackson.
\newblock Classical electrodynamics, 1999.

\bibitem[Jing et~al.(1998)Jing, Reed, and Others]{Jing1998}
J.~Jing, J.~Reed, and Others.
\newblock Automated high resolution optical mapping using arrayed, fluid-fixed
  {DNA} molecules.
\newblock \emph{Proc. Nat. Acad. Sci.}, 95\penalty0 (14):\penalty0 8046--8051,
  1998.

\bibitem[Kaplan and Mahadevan(2015)]{kaplan2015evaporation}
C.~N. Kaplan and L.~Mahadevan.
\newblock Evaporation-driven ring and film deposition from colloidal droplets.
\newblock \emph{J. Fluid Mech.}, 781, 2015.

\bibitem[Kellogg(1929)]{Kellogg1929}
O.~D. Kellogg.
\newblock \emph{Foundations of potential theory}.
\newblock Springer, 1929.

\bibitem[Layani et~al.(2009)Layani, Gruchko, Milo, Balberg, Azulay, and
  Magdassi]{Layani2009}
M.~Layani, M.~Gruchko, O.~Milo, I.~Balberg, D.~Azulay, and S.~Magdassi.
\newblock Transparent conductive coatings by printing coffee ring arrays
  obtained at room temperature.
\newblock \emph{ACS Nano}, 3\penalty0 (11):\penalty0 3537--3542, 2009.

\bibitem[Lee and Chien(1994)]{lee1994electrostatics}
C.~C. Lee and D.~H. Chien.
\newblock Electrostatics and thermostatics: a connection between electrical and
  mechanical engineering.
\newblock \emph{Int. J., Engg Ed. Vol}, 10\penalty0 (5):\penalty0 434--449,
  1994.

\bibitem[Lohse et~al.(2015)Lohse, Zhang, et~al.]{lohse2015surface}
D.~Lohse, X.~Zhang, et~al.
\newblock Surface nanobubbles and nanodroplets.
\newblock \emph{Rev. Mod. Phys.}, 87\penalty0 (3):\penalty0 981, 2015.

\bibitem[Mai and Richerzhagen(2007)]{mai200753}
T.~A. Mai and B.~Richerzhagen.
\newblock 53.3: Manufacturing of 4th generation oled masks with the laser
  microjet{\textregistered} technology.
\newblock In \emph{SID Symposium Digest of Technical Papers}, volume~38, pages
  1596--1598. Wiley Online Library, 2007.

\bibitem[Mallinson and de~Vahl~Davis(1973)]{mallinson1973method}
G.~D. Mallinson and G.~de~Vahl~Davis.
\newblock The method of the false transient for the solution of coupled
  elliptic equations.
\newblock \emph{J. Comp. Phys.}, 12\penalty0 (4):\penalty0 435--461, 1973.

\bibitem[Mampallil and Eral(2018)]{mampallil2018review}
D.~Mampallil and H.~B. Eral.
\newblock A review on suppression and utilization of the coffee-ring effect.
\newblock \emph{Adv. Coll. \& Interface Sci.}, 252:\penalty0 38--54, 2018.

\bibitem[Moore et~al.(2021)Moore, Vella, and Oliver]{moore2021nascent}
M.~R. Moore, D.~Vella, and J.~M. Oliver.
\newblock The nascent coffee ring: how solute diffusion counters advection.
\newblock \emph{J. Fluid Mech.}, 920, 2021.

\bibitem[Moore et~al.(2022)Moore, Vella, and Oliver]{moore2022nascent}
M.~R. Moore, D.~Vella, and J.~M. Oliver.
\newblock The nascent coffee ring with arbitrary droplet contact set: an
  asymptotic analysis.
\newblock \emph{J. Fluid Mech.}, 940, 2022.

\bibitem[Murisic and Kondic(2011)]{murisic2011evaporation}
N.~Murisic and L.~Kondic.
\newblock On evaporation of sessile drops with moving contact lines.
\newblock \emph{J. Fluid Mech.}, 679:\penalty0 219--246, 2011.

\bibitem[Okon and Harrington(1979)]{okon1979capacitance}
E.~E. Okon and R.~F. Harrington.
\newblock The capacitance of discs of arbitrary shape.
\newblock Technical report, Syracuse Univ. NY Dept. Elect. and Comp. Eng.,
  1979.

\bibitem[Oliver et~al.(2015)Oliver, Whiteley, Saxton, Vella, Zubkov, and
  King]{oliver2015contact}
J.~Oliver, J.~Whiteley, M.~Saxton, D.~Vella, V.~Zubkov, and J.~King.
\newblock On contact-line dynamics with mass transfer.
\newblock \emph{Eur. J. Appl. Math.}, 26\penalty0 (5):\penalty0 671--719, 2015.

\bibitem[Orejon et~al.(2011)Orejon, Sefiane, and Shanahan]{orejon2011stick}
D.~Orejon, K.~Sefiane, and M.~E. Shanahan.
\newblock Stick--slip of evaporating droplets: substrate hydrophobicity and
  nanoparticle concentration.
\newblock \emph{Langmuir}, 27\penalty0 (21):\penalty0 12834--12843, 2011.

\bibitem[Popov(2005)]{popov2005evaporative}
Y.~O. Popov.
\newblock Evaporative deposition patterns: spatial dimensions of the deposit.
\newblock \emph{Phys. Rev. E}, 71\penalty0 (3):\penalty0 036313, 2005.

\bibitem[Popov and Witten(2003)]{popov2003characteristic}
Y.~O. Popov and T.~A. Witten.
\newblock Characteristic angles in the wetting of an angular region: Deposit
  growth.
\newblock \emph{Phys. Rev. E}, 68\penalty0 (3):\penalty0 036306, 2003.

\bibitem[Rienstra(1990)]{rienstra1990shape}
S.~Rienstra.
\newblock The shape of a sessile drop for small and large surface tension.
\newblock \emph{J. Engg. Math.}, 24\penalty0 (3):\penalty0 193--202, 1990.

\bibitem[S{\'a}enz et~al.(2017)S{\'a}enz, Wray, Che, Matar, Valluri, Kim, and
  Sefiane]{saenz2017dynamics}
P.~S{\'a}enz, A.~Wray, Z.~Che, O.~Matar, P.~Valluri, J.~Kim, and K.~Sefiane.
\newblock Dynamics and universal scaling law in geometrically-controlled
  sessile drop evaporation.
\newblock \emph{Nature Comm.}, 8\penalty0 (1):\penalty0 1--9, 2017.

\bibitem[Smith and Brutin(2018)]{smith2018wetting}
F.~Smith and D.~Brutin.
\newblock Wetting and spreading of human blood: Recent advances and
  applications.
\newblock \emph{Curr. Opin. in Coll. \& Inter. Sci.}, 36:\penalty0 78--83,
  2018.

\bibitem[Sneddon(1966)]{sneddon1966mixed}
I.~N. Sneddon.
\newblock \emph{Mixed boundary value problems in potential theory}.
\newblock North-Holland Publishing Company, 1966.

\bibitem[Sultan et~al.(2005)Sultan, Boudaoud, and Amar]{sultan2005evaporation}
E.~Sultan, A.~Boudaoud, and M.~B. Amar.
\newblock Evaporation of a thin film: diffusion of the vapour and marangoni
  instabilities.
\newblock \emph{J. Fluid Mech.}, 543:\penalty0 183--202, 2005.

\bibitem[Van~Dyke(1964)]{VanDyke1964}
M.~Van~Dyke.
\newblock \emph{Perturbation methods in fluid mechanics}.
\newblock Academic Press New York, 1964.

\bibitem[Weber(1873)]{weber1873ueber}
H.~Weber.
\newblock {\"U}ber die {B}esselschen {F}unctionen und ihre {A}nwendung auf die
  {T}heorie der elektrischen {S}tr{\"o}me.
\newblock \emph{J. f{\"u}r die reine Angew. Math.}, 75:\penalty0 75--105, 1873.

\bibitem[Weon and Je(2013)]{weon2013self}
B.~M. Weon and J.~H. Je.
\newblock Self-pinning by colloids confined at a contact line.
\newblock \emph{Phys. Rev. Lett.}, 110\penalty0 (2):\penalty0 028303, 2013.

\bibitem[Witten(2009)]{witten2009robust}
T.~Witten.
\newblock Robust fadeout profile of an evaporation stain.
\newblock \emph{EPL}, 86\penalty0 (6):\penalty0 64002, 2009.

\bibitem[Wray et~al.(2014)Wray, Papageorgiou, Craster, Sefiane, and
  Matar]{wray2014electrostatic}
A.~W. Wray, D.~T. Papageorgiou, R.~V. Craster, K.~Sefiane, and O.~K. Matar.
\newblock Electrostatic suppression of the “coffee stain effect”.
\newblock \emph{Langmuir}, 30\penalty0 (20):\penalty0 5849--5858, 2014.

\bibitem[Wray et~al.(2021)Wray, Wray, Duffy, and Wilson]{wray2021contact}
A.~W. Wray, P.~S. Wray, B.~R. Duffy, and S.~K. Wilson.
\newblock Contact-line deposits from multiple evaporating droplets.
\newblock \emph{Phys. Rev. Fluids}, 6\penalty0 (7):\penalty0 073604, 2021.

\bibitem[Zheng et~al.(2005)Zheng, Popov, and Witten]{zheng2005deposit}
R.~Zheng, Y.~O. Popov, and T.~A. Witten.
\newblock Deposit growth in the wetting of an angular region with uniform
  evaporation.
\newblock \emph{Phys. Rev. E}, 72\penalty0 (4):\penalty0 046303, 2005.

\end{thebibliography}

\end{document}